\documentclass[letterpaper]{IEEEtran}
\usepackage{amsmath}
\usepackage{amssymb, amsbsy,stmaryrd}
\usepackage{tabularx,comment}
\usepackage{graphics,graphicx}
\usepackage[usenames]{color}
\usepackage{fixltx2e}  % prevents starred figures to be placed out of order

\newcommand{\squeeze}[1][{}]{}

\newcommand{\iverson}[1]{\mathbf{1}_{#1}}

\newcommand{\alp}{\mathcal{A}}
\newcommand{\alpb}{\hat{\alp}}
\newcommand{\Ak}{\alpb_k}
\newcommand{\Akprime}{\Ak^\ast}

\newcommand{\Cplus}[1][k]{C_{#1}}
\newcommand{\Cminus}[1][k]{C_{-{#1}}}
\newcommand{\limcode}{\Cminus[\infty]}
\newcommand{\limtree}{\limcode}

\newcommand{\depth}[2][T]{\textrm{depth}_{#1}(#2)}

\newcommand{\optq}[1][q]{\mathcal{T}_{#1}}

\newcommand{\fringeT}[1][T]{f_{#1}}
\newcommand{\Tsc}[1][\short,c]{T_{#1}}
\newcommand{\Tc}[1][c]{T_{\short,#1}}
\newcommand{\Tck}{\Tc[{c_k}]}
\newcommand{\Tskck}{\Tsc[{\short_k,\,c_k}]}
\newcommand{\Tsgcg}{\Tsc[{\short_g,\,c_g}]}
\newcommand{\fT}[1][T]{f_{#1}}
\newcommand{\nT}[1][T]{\mathbf{N}_{#1}}
\newcommand{\nN}{\mathbf{N}}
\newcommand{\mcT}{\mathcal{T}}
\newcommand{\Tg}[2][g]{\mathcal{T}_{#1}^{#2}}

\newcommand{\weight}[2][{}]{w_{#1}(#2)}
\newcommand{\wt}{w}

\newcommand{\costt}{\mathcal{L}}
\newcommand{\cost}[2][q]{\costt_{#1}(#2)}
\newcommand{\ncostt}{\overline{\mathcal{L}}}
\newcommand{\ncost}[2][q]{\ncostt_{#1}(#2)}

\newcommand{\Lc}[1][c]{L_{\short,#1}}
\newcommand{\limlen}{\ncost{\limcode}}

\newcommand{\Prob}{P}

\newcommand{\rsymb}[1][s]{\mathcal{R}_{#1}}

\newcommand{\tdgd}{\mbox{TDGD}}
\newcommand{\tdgds}{{\tdgd}s}
\newcommand{\eq}[1]{(\ref{#1})}
\newcommand{\qinv}{q^{-1}}
\newcommand{\eps}{\varepsilon}
\newcommand{\qeps}{q_{1}}

\newcommand{\half}{{\frac{1}{2}}}
\newcommand{\eighth}{{\frac{1}{8}}}

\newcommand{\bldp}{\mathbf{p}}

\newcommand{\ccmin}[1][\short]{{\underline{c}_{\,#1}}}  % min possible value of c
\newcommand{\ccmax}[1][\short]{{\overline{c}_{#1}}}  % max possible value of c
\newcommand{\len}[1]{\left\lvert{#1}\right\rvert}
\newcommand{\profileT}[1][T]{n^{#1}}
\newcommand{\qparI}[1][k]{2^{-#1}}
\newcommand{\qparII}[1][k]{2^{\,-1/#1}}
\newcommand{\sg}{\text{sg}}

\renewcommand{\paragraph}[1]{\noindent\textbf{#1}}

\newtheorem{lemma}{Lemma}
\newtheorem{corollary}{Corollary}
\newtheorem{theorem}{Theorem} %[section]

\newtheorem{EXAMPLE}{Example}[section]

\newcommand{\concat}{\cdot}
\renewcommand{\boxed}[1]{{\setlength\fboxsep{1.5pt}\fbox{${#1}$}}}

\newcommand{\macrosymbol}{\boldsymbol{\mathcal{M}}}
\newcommand{\layer}[1][s]{\mathbf{L}_{#1}}
\newcommand{\defined}{\stackrel{\scriptscriptstyle\Delta}{=}}
\newcommand{\lambdak}[1][k]{\lambda_{#1}}
\newcommand{\Mk}[1][k]{M_{#1}}

\newcommand{\short}{\sigma}
\newcommand{\shortm}{\short_{\!\scriptscriptstyle -}}
\newcommand{\shortp}{\short_{\!\scriptscriptstyle +}}
\newcommand{\cmn}{c_{\scriptscriptstyle -}}
\newcommand{\cpl}{c_{\scriptscriptstyle +}}
\newcommand{\DD}[1][c]{D_{{\short,{#1}}}}
\newcommand{\DDsc}[1][{\short,c}]{D_{{#1}}}
\newcommand{\symb}{\psi}
\newcommand{\cerouno}{\{0,1\}}
\newcommand{\sseq}{\mathbf{s}}

\newcommand{\pair}{(\short, c)}
\newcommand{\cck}{c_k}

\newcommand{\pairk}{(\short_k,c_k)}
\newcommand{\Qk}{Q}
\newcommand{\bldeps}{{\boldsymbol{\eps\!\!\!\eps}}}

\newcommand{\fxdelta}{\Delta}
\newcommand{\Ffxn}{F}
\newcommand{\predicate}{\mathcal{P}}
\newcommand{\jtuple}{j,r,j',r'}
\newcommand{\jvector}{\mathbf{j}}
\newcommand{\epstuple}{\eps,\eps'{\!\!,\,}\eps''{\!\!\!,\:}\eps'''}
\newcommand{\subroot}{\nu}

\newcommand{\Kset}{\mathcal{K}}
\newlength{\saveunitlength}
\newcommand{\tree}[3][-3ex]{%
\setlength{\saveunitlength}{\unitlength}
{\raisebox{#1}{
\setlength{\unitlength}{0.05em}
\begin{picture}(45,40)(0,-5)
\put(20,50){\circle*{6}}
\put(20,50){\line(-1,-3){10}}
\put(20,50){\line(1,-3){10}}
\put(5,14){\makebox(0,0)[t]{#2}}
\put(35,14){\makebox(0,0)[t]{#3}}
\end{picture}
\setlength{\unitlength}{\saveunitlength}
}}}

\newcommand{\btree}[3][{}]{%
\setlength{\saveunitlength}{\unitlength}
\setlength{\unitlength}{0.05em}
{\raisebox{1.5em}{
\begin{picture}(65,60)(0,30)
\put(30,50){\circle*{6}}
\put(40,60){\makebox(0,0)[l]{#1}}
\put(30,50){\line(-2,-3){20}}
\put(30,50){\line(2,-3){20}}
\put(10,12){\makebox(0,0)[t]{#2}}
\put(50,12){\makebox(0,0)[t]{#3}}
\end{picture}
\setlength{\unitlength}{\saveunitlength}
}}}

\newcommand{\stackT}[2][-1.5ex]%
{\underbrace{\rule[#1]{0pt}{0pt}{\macrosymbol{\,\ldots\,}\macrosymbol}}%
_{#2\;\text{times}}}
\newcommand{\treerep}[4][-3ex]{%
\underbrace{\rule[-3.5ex]{0pt}{0pt}{\tree[#1]{#2}{#3}\!\!\!{\ldots}\!\!\!\tree[#1]{#2}{#3}}}%
_{\scriptstyle#4\;\text{times}}}
\newcommand{\leafrep}[2][-1.5ex]{%
\underbrace{\rule[#1]{0pt}{0pt}\boxed{1}{\,\ldots\,}\boxed{1}}%
_{\scriptstyle #2\;\text{times}}}

\newcommand{\Ctree}{\mathcal{U}}
\newcommand{\Ltree}{\mathcal{V}}
\newcommand{\Ltreek}[1][k]{\Ltree_{#1}}
\newcommand{\Ltreem}[1][k]{\Ltreek[#1]^{-}}

\newcommand{\FFF}{V}
\newcommand{\FFFP}{V^\ast}
\newcommand{\DDD}{D}

\title{Optimal prefix codes for pairs of geometrically-distributed
random variables}

\author{
\IEEEauthorblockN{Fr\'ed\'erique Bassino, %
Julien Cl\'ement, %
Gadiel Seroussi, \IEEEmembership{Fellow, IEEE,} %
and Alfredo Viola }

\thanks{
This work was supported in part by ECOS project U08E02,  by
PDT project 54/178 2006--2008, and by CSIC project
(Universidad de la Rep\'{u}blica) fondos 2009--2011.
A. Viola's work was done in part while he
was visiting GREYC, Universit\'{e} de Caen and
the Laboratoire d'Informatique Gaspard-Monge,
Universit\'{e} de Marne la Vall\'{e}e, France. %
Parts of this paper were presented
at the 2006 Data Compression Conference, and at the
2006 IEEE International Symposium on Information Theory.

F. Bassino is with LIPN UMR 7030.
Universit\'e Paris 13 - CNRS, France (e-mail:
bassino@lipn.univ-paris13.fr).

J. Cl\'ement is with
GREYC UMR 6072, CNRS, Universit\'e de Caen, ENSICAEN, France
(e-mail: julien.clement@unicaen.fr).

G. Seroussi is with Hewlett-Packard Laboratories, Palo Alto, CA 94304,
USA, and with Facultad de Ingenier\'{\i}a,
Universidad de la Rep\'ublica, Montevideo, Uruguay (e-mail:gseroussi@ieee.org).

A. Viola is with
Instituto de Computaci\'{o}n, Facultad de Ingenier\'{\i}a, %
Universidad de la Rep\'ublica, Montevideo, Uruguay %
(e-mail: viola@fing.edu.uy).

Copyright (c) 2012 IEEE. Personal use of this material is permitted.
However, permission to use this material for any other purposes must be
obtained from the IEEE by sending a request to
pubs-permissions@ieee.org. }}

\begin{document}
\maketitle

\begin{abstract}
Optimal prefix codes are studied for pairs of independent,
integer-valued symbols emitted by a source with a geometric probability
distribution of parameter $q$, $0{<}q{<}1$. By encoding pairs of
symbols, it may be possible to reduce the redundancy penalty of
symbol-by-symbol encoding, while preserving the simplicity of the
encoding and decoding procedures typical of Golomb codes and their
variants. It is shown that optimal codes for these so-called
two-dimensional geometric distributions are \emph{parameter-singular},
in the sense that a prefix code that is optimal for one value of the
parameter $q$ cannot be optimal for any other value of $q$. This is in
sharp contrast to the one-dimensional case, where codes are optimal for
positive-length intervals of the parameter $q$. Thus, in the
two-dimensional case, it is infeasible to give a compact
characterization of optimal codes for all values of the parameter $q$,
as was done in the one-dimensional case. Instead, optimal codes are
characterized for a discrete sequence of values of $q$ that provides
good coverage of the unit interval. Specifically, optimal prefix codes
are described for $q=\qparII$ ($k\ge 1$), covering the range $q\ge
\half$, and $q=\qparI$ ($k>1$), covering the range $q<\half$. The
described codes produce the expected reduction in redundancy with
respect to the one-dimensional case, while maintaining low complexity
coding operations.

\medskip

\end{abstract}

\section{Introduction}\label{sec:intro}
In 1966, Golomb~\cite{gol66} described optimal binary prefix codes for
some geometric distributions over the nonnegative integers, namely,
distributions with probabilities $p(i)$ of the form
\[
p(i) = (1-q)q^i \,,\quad i\geq 0,
\]
for some real-valued parameter $q$, $0 < q<1$. In~\cite{gv75}, these
\emph{Golomb codes} were shown to be optimal for \emph{all} geometric
distributions. These distributions occur, for example, when encoding
\emph{run lengths} (the original motivation in~\cite{gol66}), and in
image compression when encoding prediction residuals, which are
well-modeled by~\emph{two-sided geometric distributions}. Optimal codes
for the latter were characterized in~\cite{msw00}, based on some
combinations and variants of Golomb codes. Codes based on the Golomb
construction have the practical advantage of allowing the encoding of a
symbol $i$ using a simple explicit computation on the integer value of
$i$, without recourse to nontrivial data structures or tables. This has
led to their adoption in many practical applications
(cf.~\cite{rice79},\cite{jpegls}).

Symbol-by-symbol encoding, however, can incur
significant redundancy relative to the entropy of the distribution,
even when dealing with sequences of independent, identically
distributed random variables. One way to mitigate this problem, while
keeping the simplicity and low latency of the encoding and decoding
operations, is to consider short blocks of $d{>}1$ symbols, and use a
prefix code for the blocks. In this paper, we study optimal prefix
codes for pairs (blocks of length $d{=}2$) of independent, identically
distributed geometric random variables, namely, distributions on pairs
of nonnegative integers $(i,j)$ with probabilities of the form
\begin{equation}\label{eq:tdgd}
\Prob(i,j) = p(i)p(j) = (1-q)^2 q^{i+j}\quad i,j\geq 0.
\end{equation}
We refer to this distribution as a \emph{two-dimensional geometric
distribution (TDGD)}, defined on the alphabet of integer pairs
$\alp=\{\,(i,j)\;|\;i,j\geq 0\,\}$. For succinctness, we denote a TDGD
of parameter $q$ by $\tdgd(q)$.

Aside from the mentioned practical motivation, the problem is of
intrinsic combinatorial interest. It was proved in \cite{ltz97} (see
also~\cite{KatoHanNagaoka96}) that, if the entropy\footnote{%
$\log x$ and $\ln x$ will denote, respectively, the base-$2$ and the
natural logarithm of $x$.}
 $-\sum_{ a\in\mathcal{A}} P(a) \log P(a)$ of a distribution over a countable
 alphabet $\mathcal{A}$ is finite, optimal codes exist and can be
obtained, in the limit, from Huffman codes for truncated versions of
the alphabet. However, the proof does not give a general way for
effectively constructing optimal codes, and in fact, there are few
families of distributions over countable alphabets for which an
effective construction is known~\cite{abr01}\cite{GolinTR04}. An
algorithmic approach to building optimal codes is presented
in~\cite{GolinTR04}, which covers geometric distributions and various
generalizations. The approach, though, is not applicable to TDGDs, as
explicitly noted in~\cite{GolinTR04}.

Some characteristic properties of the families of optimal codes for
geometric and related distributions in the one-dimensional case turn
out not to hold in the two-dimensional case. Specifically, the optimal
codes described in~\cite{gol66} and~\cite{msw00} correspond to binary
trees of \emph{bounded width}, namely, the number of codewords of any
given length is upper-bounded by a quantity that depends only on the
code parameters. Also, the family of optimal codes in each case
partitions the parameter space into regions of positive volume, such
that all the corresponding distributions in a region admit the same
optimal code. These properties do not hold in the case of optimal codes
for TDGDs. In particular, optimal codes for TDGDs turn out to be
\emph{parameter-singular}, in the sense that if a code $\mathcal{T}_q$ is optimal
for $\tdgd(q)$, then $\mathcal{T}_q$ \emph{is not} optimal for
$\tdgd(q')$ for any parameter value $q'\ne q$. This result is presented
in Section~\ref{sec:singular}.
 (A related but somewhat dual problem, namely, counting the number
 of distinct trees that can be optimal for a given source over a
 countable alphabet, is studied in~\cite{Gol:80}.)

An important consequence of this singularity is that any set containing
optimal codes for all values of $q$ must be uncountable, and, thus, it
would be infeasible to give a compact characterization of such a set,
as was done in~\cite{gol66} or \cite{msw00} for  one-dimensional
cases.\footnote{Loosely, by a compact characterization we mean one in
which each code is characterized by a finite number of finite
parameters, which drive the corresponding encoding/decoding
procedures.} Thus, from a practical point of view, the best we can
expect is to characterize optimal codes for countable sequences of
parameter values. In this paper, we present such a characterization,
for a sequence of parameter values that provides good coverage of the
range of $0{<}q{<}1$. Specifically, in Section~\ref{sec:q>=1/2}, we
describe the construction of optimal codes for $\tdgd(q)$ with
$q=\qparII$ for integers $k\ge 1$,%
\footnote{ These are the same distributions for which optimality of
Golomb codes was originally established in~\cite{gol66}.} %
covering the range $q\ge \half$, and in Section~\ref{sec:q<1/2}, we do
so for $\tdgd(q)$ with $q=\qparI$ for integers $k>1$, covering the
range $q<\half$ (thus, overall, we show optimal codes for all values of
$q$ such that $-\log q$ is either an integer or the inverse of one). In
the case $q<\half$, we observe that, as  $k\to\infty$ ($q\to 0$), the
optimal codes described converge to a  \emph{limit code}, in the sense
that the codeword for any given pair  $(a,b)$ remains the same for all
$k > k_0(a,b)$, where $k_0$ is a  threshold that can be computed from
$a$ and $b$ (this limit code is also mentioned, without proofs,
in~\cite{MBaer2003}). The codes in both constructions are of unbounded
width. However, they are
\emph{regular}~\cite{regular_trees}, in the sense that the
corresponding infinite trees have only a finite number of
non-isomorphic \emph{whole subtrees} (i.e., subtrees consisting of a
node and all of its descendants). This allows for deriving  recursions
and  explicit expressions for the average code length,  as well as
feasible encoding/decoding procedures. Notice that, to the best of our
knowledge, the only case for which an optimal code for a TDGD had been
characterized prior to this work was the trivial case $q=\half$, in
which case encoding each component of $(i,j)$ separately with a unary
code (i.e., a Golomb code of order one) has zero redundancy, and is
thus optimal (cf.\ also~\cite{MBaer2003}).

Practical considerations, and the redundancy of the new codes, are
discussed in Section~\ref{sec:plots}, where we present redundancy plots
and comparisons with  symbol-by-symbol Golomb coding and with the
optimal code for a TDGD for each plotted value of $q$ (optimal average
code lengths for arbitrary values of $q$ were estimated numerically to
sufficiently high precision). We also derive an exact expression for
the asymptotic oscillatory behavior of the redundancy of the new codes
as $q\to 1$. The study confirms the redundancy gains over
symbol-by-symbol encoding with Golomb codes, and the fact that the
discrete sequence of codes presented provides a good approximation to
the full class of optimal codes over the range of the parameter $q$.

Our constructions and proofs of optimality rely on the technique of
Gallager and Van Voorhis~\cite{gv75}, which was also used
in~\cite{msw00}. As noted in~\cite{gv75}, most of the work and
ingenuity in applying the technique goes into discovering appropriate
``guesses'' of the basic components on which the construction iterates,
and in describing the structure of the resulting codes. With the
correct guesses, the proofs are straightforward. The technique
of~\cite{gv75} is reviewed in Section~\ref{sec:prelim}, where we also
introduce some definitions and notation that will be useful throughout
the paper.

\section{Preliminaries}\label{sec:prelim}

\subsection{Definitions}\label{sec:defs}

We are interested in encoding the alphabet $\alp$ of integer pairs $(i,j)$,
$i,j\geq 0$, using a binary prefix code $C$ (we will refer to $C$ plainly as
a \emph{code}, the binary and prefix properties assumed throughout). As
usual, we associate $C$ with a rooted (infinite) binary tree, whose leaves
correspond, bijectively, to symbols in $\alp$, and where each branch is
labeled with a binary digit. The binary codeword assigned to a symbol is
``read off'' the labels on the path from the root to the corresponding leaf.
The \emph{depth} of a node $x$ in a tree $T$, denoted $\depth[T]{x}$, is the
number of branches on the path from the root to $x$.  By extension, the depth
(or \emph{height}) of a finite tree is defined as the maximal depth of any of
its nodes. A \emph{level} of $T$ is the set of all nodes at a given depth
$\ell$ (we refer to this set as \emph{level} $\ell$). Let $\profileT_{\ell}$
denote the number of leaves in level $\ell$ of $T$ (we will sometimes omit
the superscript $T$ when clear from the context). We refer to the sequence
$\{\profileT_{\ell}\}_{\ell\ge 0}$ as the \emph{profile} of $T$.  Two trees
will be considered \emph{equivalent} if their profiles are identical. Thus,
for a code $C$, we are only interested in its tree profile, or, equivalently,
the \emph{length distribution} of its codewords. Given the profile of a tree,
and an ordering of $\alp$ in decreasing probability order, it is always
possible to define a canonical tree (say, by assigning leaves in alphabetical
order; see, e.g.,~\cite{Cover2ndEdition}) that uniquely defines a code for
$\alp$. The notion of tree equivalence adopted implies that given a tree, we
can arbitrarily permute the nodes at any level, since such a permutation
leaves the profile invariant. This will allow us to make, without loss of
generality, certain assumptions on the structure of the tree. In particular,
we will often make the assumption that if a tree contains, say, at least
$2^j$ leaves at a certain level $\ell$, then there is a set of $2^j$ leaves
at level $\ell$ that have a common ancestor\footnote{We use the usual
``family'' terminology for trees: nodes have children, parents, ancestors and
descendants. We also use the common convention of visualizing trees with the
root at the top and leaves at the bottom. Thus, ancestors are ``up,'' and
descendants are ``down.''} $\nu$ at level $\ell-j$ (an alphabetically ordered
tree, in fact, always has this property).

With a slight abuse of terminology, we will not distinguish between a
code and its corresponding tree (or profile), and will refer to the
same object sometimes as a tree and sometimes as a code. Unless noted
otherwise, all trees considered in this paper are \emph{full}, i.e.,
every node in the tree is either a leaf or the parent of two children
(full trees are sometimes referred to in the literature as
\emph{complete}). A tree is \emph{balanced} (or
\emph{uniform}) if it has $2^k$ leaves, all of them at depth $k$,
for some $k\ge0$. We denote such a tree by $\Ctree_k$. We will restrict
the use of the term \emph{subtree} to refer to whole subtrees of $T$,
i.e., subtrees that consist of a node and all of its descendants in
$T$.

We call $s(i,j)=i+j$ the \emph{signature} of $(i,j)\in\alp$. For a
given value $s=s(i,j)$, there are $s{+}1$ pairs with signature $s$, all
with the same probability, $P(s){=}(1-q)^2 q^s$, under the
distribution~(\ref{eq:tdgd}). Given a code $C$, symbols of the same
signature can be freely permuted without affecting the properties of
interest to us (e.g., average code length). Thus, for simplicity, we
can also regard the correspondence between leaves and symbols as one
between leaves and elements of the \emph{multiset}
\begin{equation}\label{eq:alpb} \alpb = \{0, 1, 1, 2,2,2,\dotsc,
\underbrace{s, \dotsc, s}_{s+1\;\text{times}}, \dotsc\}.
\end{equation}
In constructing the tree, we do not distinguish between different
occurrences of a signature $s$; for actual encoding, the $s{+}1$ leaves
labeled with $s$ are mapped to the symbols
$(0,s),(1,s{-}1),\ldots,(s,0)$ in some fixed order. In the sequel, we
will often ignore normalization factors for the signature probabilities
$P(s)$ (in cases where normalization is inconsequential), and will
use instead \emph{weights} $w(s)=q^s$.

Consider a tree (or code) $T$ for $\alp$. Let $U$ be a subtree of $T$,
and let $s(x)$ denote the signature associated with a leaf $x$ of $U$.
Let $F(U)$ denote the set of leaves of $U$, referred to as its
\emph{fringe}. We define the
\emph{weight\/}, $\weight[q]{U}$, of $U$ as
\[
\weight[q]{U}  = \sum_{x\in F(U)}
q^{s(x)}\,,
\]
 and the
\emph{cost\/}, $\cost{U}$,  of $U$
 as
\[
\cost{U}  =
\sum_{x\in F(U)} \depth[U]{x} q^{s(x)}\,
\]
(the subscript $q$ may be omitted when clear from the context). When
$U=T$, we have $\weight[q]{T}=(1-q)^{-2}$, and $\ncost{T} \defined
(1-q)^2\cost{T}$ is the average code length of $T$. A tree $T$ is
\emph{optimal} for $\tdgd(q)$ if $\cost{T} \le \cost{T'}$ for any tree
$T'$.

\subsection{Some basic objects and operations}
\label{sec:basics}
For $\alpha \geq 1$, we say that a finite source with probabilities
$p_1 \ge p_2 \ge \dotsm \ge p_{N}$, $N\ge 2$, is {\em $\alpha$-uniform}
if $p_1/p_N\leq \alpha$. A $2$-uniform source is also called {\em
quasi-uniform}. An optimal code for a quasi-uniform source on $N$
symbols consists of $2^{\lceil \log N \rceil}{-}N$ codewords of length
$\lfloor
\log N \rfloor$, and $2N{-}2^{\lceil \log N \rceil}$ codewords of
length $\lceil \log N \rceil$, the shorter codewords corresponding to
the more probable symbols~\cite{gv75}. We refer to such a code (or the
associated tree) also as \emph{quasi-uniform}, denote it by $Q_N$, and
denote by $Q_N(i)$ the codeword it assigns to the symbol associated
with $p_i$, $1{\le}i{\le}N$. For convenience, we define $Q_1$ as a null
code, which assigns code length zero to the single symbol in the
alphabet. Clearly, for integers $k\ge 0$, we have $Q_{2^k}=\Ctree_k$.
The
\emph{fringe thickness} of a finite tree $T$, denoted $\fringeT$, is
the maximum difference between the depths of any two leaves of $T$.
Quasi-uniform trees $T$ have $\fringeT\leq 1$, while uniform trees have
$\fringeT=0$. In Section~\ref{sec:q>=1/2}  we present a
characterization of optimal codes of fringe thickness two for
$4$-uniform distributions, which generalizes the quasi-uniform case.
This generalization will help in the characterization of the optimal
codes for $\tdgd(q)$, $q=\qparII$.

The \emph{concatenation} of two trees $T$ and $U$, denoted $T\concat
U$, is obtained by attaching a copy of $U$ to each leaf of $T$.
Regarded as a code, $T\concat U$ consists of all the possible
concatenations $t\concat u$ of a word $t\in T$ with one $u\in U$. The
\emph{Golomb code} of order $k \ge 1$~\cite{gol66}, denoted $G_k$,
encodes an integer $i$ by concatenating $Q_k(i \bmod k)$ with a
\emph{unary} encoding of $\lfloor i/k \rfloor$ (e.g., $\lfloor i/k
\rfloor$ ones followed by a zero). The first-order Golomb code $G_1$ is
just the unary code, whose corresponding tree consists of a root with
one leaf child on the branch labeled '0', and, recursively, a copy of
$G_1$ attached to the child on the branch labeled '1'. Thus, we have
$G_k = Q_k \concat G_1$.

\subsection{The Gallager-Van Voorhis method}\label{sec:gv}

When proving optimality of infinite codes for TDGDs, we will rely on the
method due to Gallager and Van Voorhis~\cite{gv75}, which is briefly outlined
below, adapted to our setting and terminology.
\begin{itemize}
\item Define a sequence of finite \emph{reduced sources}
    $(\mathcal{S}_{t})_{t=0}^\infty$. The alphabet of the reduced source
    $\mathcal{S}_t$ is a multiset $\mathcal{S}_t = \mathcal{H}_t \cup
    \mathcal{F}_t$, where $\mathcal{H}_t$ is a multiset comprising the
    signatures $0,1,\ldots,s{-}1$ (with multiplicities as
    in~(\ref{eq:alpb})), and $\mathcal{F}_t$ consists of a finite number of
    (possibly infinite) subsets of $\alpb$, referred to as \emph{virtual
    symbols}, which form a partition of the remaining signatures. We
    naturally associate with each virtual symbol a weight equal to the sum
    of the weights of the signatures it contains.
\item Verify that the sequence $(\mathcal{S}_{t})_{t=0}^\infty$ is
    compatible with the bottom-up Huffman procedure. This means that after
    a number of merging steps of the Huffman algorithm on the reduced
    source $\mathcal{S}_t$, one gets $\mathcal{S}_{t-1}$. Proceed
    recursively, until $\mathcal{S}_0$ is obtained.
\item Apply the Huffman algorithm to $\mathcal{S}_{0}$.
\end{itemize}

While the sequence of reduced sources $\mathcal{S}_{t}$ can be seen as
evolving ``bottom-up,'' the infinite code $C$ constructed results from a
``top-down'' sequence of corresponding finite codes $C_{t}$, whose size grows
with $t$, and which unfold by recursive reversal of the mergers in the Huffman
procedure. One shows that the sequence of codes $(C_{t})_{t\geq 0}$ {\em
converges} to an infinite code $C$, in the sense that for every $j\geq 1$,
with codewords of $C_{t}$ consistently sorted, the $j$th codeword of $C_{t}$
is eventually constant when $t$ grows, and equal to the $j$th codeword of
$C$. A corresponding convergence argument on the sequence of average code
lengths then establishes the optimality of $C$.

This method was successfully applied to characterize infinite optimal
codes in~\cite{gv75} and~\cite{msw00}. While the technique is
straightforward once appropriate reduced sources are defined, the
difficulty in each case is to guess the structure of these source. In a
sense, this is a self-bootstrapping procedure, where one needs to guess
the structure of the codes sought, and use that structure to define the
reduced sources, which, in turn, serve to prove that the guess was
correct. We will apply the Gallager-Van Voorhis method to prove
optimality of codes for certain families of TDGDs in
Sections~\ref{sec:q>=1/2} and~\ref{sec:q<1/2}. In each case, we will
emphasize the definition and structure of the reduced sources, and show
that they are compatible with the Huffman procedure. We will omit the
discussion on convergence, and the formal induction proofs, since the
arguments are essentially the same as those in~\cite{gv75}
and~\cite{msw00}.

\section{Parameter-singularity of optimal codes for TDGDs}
\label{sec:singular}

In the case of one-dimensional geometric distributions, the unit
interval $(0,1)$ is partitioned into an infinite sequence of semi-open
intervals $(q_{k-1},q_{k}]$, $k\ge 1$, such that the Golomb code $G_k$
is optimal for all values of the distribution parameter $q$ in
$(q_{k-1},q_{k}]$. Specifically, for $k\ge 0$, $q_k$ is the (unique)
nonnegative root of the equation $q^{k}+q^{k+1}-1 = 0$~\cite{gv75}.
Thus, we have $q_0=0,\,q_1=(\sqrt{5}-1)/2\approx 0.618, q_2\approx
0.755$, etc. A similar property holds in the case of two-sided
geometric distributions~\cite{msw00}, where the two-dimensional
parameter space is partitioned into a countable sequence of patches
such that all the distributions with parameter values in a given patch
admit the same optimal code. In this section, we prove that, in sharp
contrast to these examples, optimal codes for TDGDs are
parameter-singular, in the sense that a code that is optimal for a
certain value of the parameter $q$ \emph{cannot} be optimal for any
other value of $q$. More formally, we present the following result.

\begin{theorem}\label{th:singular} Let $q$ and $\qeps$ be real
numbers in the interval $(0,1)$, with $q\ne \qeps$, and let $\optq$ be
an optimal tree for $\tdgd(q)$. Then, $\optq$ is not optimal for
$\tdgd(\qeps)$.
\end{theorem}

\smallskip

\textbf{Remark.} It follows from Theorem~\ref{th:singular}
that any set containing an optimal code for each distribution
$\tdgd(q)$, for all values of $q$, must be uncountable. This implies,
in turn, that most optimal codes for TDGDs do not have finite
descriptions, in sharp contrast with the one-dimensional case. From an
algorithmic point of view, then, the key question is for what
``interesting''
\emph{countable} sets of values of $q$ a full characterization
of optimal codes is possible. In a theoretical sense, perhaps the
ultimate such set would be that of all values of $q$ which have finite
descriptions (more formally, the set of \emph{computable} values of $q$
relative to some universal Turing machine; see,
e.g.,~\cite{GareyJohnson79}). For this set, the goal would be to obtain
a general procedure which, given a finite description of $q$, and a
pair $(i,j)$, produces the corresponding codeword in an optimal code
for $\tdgd(q)$. A somewhat less ambitious theoretical goal, although
probably not less valuable from a practical point of view, would be to
characterize optimal codes for a dense countable set of values of $q$,
e.g., all rational values of $q$, or all values of $q$ such that $\log
q$ is rational. These comprehensive characterizations appear quite
challenging, and remain open problems. In Sections~\ref{sec:q>=1/2}
and~\ref{sec:q<1/2} we characterize optimal codes for a ``smaller''
infinite countable set of TDGDs, namely, the set of distributions
$\tdgd(q)$ such that $-\log q$ is either a positive integer or the
inverse of one. It will turn out, as will be shown in
Section~\ref{sec:plots}, that this set provides good coverage of the
interval $0 < q < 1$, in the sense that, given an arbitrary value $q'$
in the interval, encoding $\tdgd(q')$ with the best available code from
the characterized set results in relatively low added redundancy, and
yields the expected redundancy gains over optimal symbol-by-symbol
encoding with Golomb codes.

We will prove Theorem~\ref{th:singular} through a series of lemmas,
which will shed more light on the structure of optimal trees for TDGDs.
For simplicity, we assume throughout that a fixed optimal tree $\optq$
is given (for a given value of $q$).

\begin{lemma}\label{lem:2levels}
Leaves with a given signature $s$ are found in at most two consecutive
levels of $\optq$.
\end{lemma}
\begin{IEEEproof}
Let $d_0$ and $d_1$ denote, respectively, the minimum and maximum
depths of a leaf with signature $s$ in $\optq$. Assume, contrary to the
claim of the lemma, that $d_1 > d_0+1$. We transform $\optq$ into a
tree $\optq'$ as follows. Pick a leaf with signature $s$ at level
$d_0$, and one at level $d_1$. Place both signatures $s$ as children of
the leaf at level $d_0$, which becomes an internal node. Pick any
signature $s'$ from a level strictly deeper than $d_1$, and move it to
the vacant leaf at level $d_1$. Tracking changes in the code lengths
corresponding to the affected signatures, and their effect on the cost,
we have
\begin{equation}\label{eq:optq'}
\cost{\optq'} = \cost{\optq} + q^s(d_0-d_1+2) - q^{s'}\delta,
\end{equation}
where $\delta$ is a positive integer. By our assumption, the quantity
multiplying $q^s$ in~(\ref{eq:optq'}) is non-positive, and we have
$\cost{\optq'} < \cost{\optq}$, contradicting the optimality of
$\optq$. Therefore, we must have $d_1 \le d_0+1$.
\end{IEEEproof}

A \emph{gap} in a tree $T$ is a non-empty set of consecutive levels
containing only internal nodes of $T$, and such that both the level
immediately above the set (assuming the set does not include level 0)
and the level immediately below it contain at least one leaf each. The
corresponding
\emph{gap size} is defined as the number of levels in the gap. It
follows immediately from Lemma~\ref{lem:2levels} that in an optimal
tree, if the largest signature above a gap is $s$, then the smallest
signature below the gap is $s+1$.

\begin{lemma}\label{lem:gaps}
Let $k = 1+\lfloor\log q^{-1}\rfloor$. Then, for all sufficiently large
$s$, the size $g$ of any gap between leaves of signature $s$ and leaves
of signature $s+1$ in $\optq$ satisfies $g \leq k-1$.
\end{lemma}
\begin{IEEEproof} We consider the cases $q>\half$, $q=\half$, and
$q<\half$ separately.

\noindent\emph{Case $q>\half$.} In this case, we have $k=1$, and the
claim of the lemma means that there can be \emph{no gaps} in the tree
from a certain level on. Assume that there is a gap between level $d$
with signatures $s$, and level $d'$ with signatures $s+1$, $d'-d\geq
2$. By Lemma~\ref{lem:2levels}, all signatures $s+1$ are either in
level $d'$ or in level $d'+1$. Without loss of generality, we can
assume that there is a subtree of $\optq$ of height at most two, rooted
at a node $v$ of depth $d'-1
\geq d+1$, and containing at least two leaves of signature $s+1$.
Hence, the weight of the subtree satisfies
\begin{equation*}%\label{wt1}
\wt(v) \geq 2q^{s+1} > q^s\,,
\end{equation*}
and switching a leaf $s$ on level $d$ with node $v$ on level $d'-1$
decreases the cost of $\optq$, in contradiction with its optimality
(when switching nodes, we carry also any subtrees rooted at them).
Therefore, there can be no gap between the level containing signatures
$s$ and $s+1$, as claimed. Notice that this holds for all values of
$s$, regardless of level.

\noindent\emph{Case $q=\half$.} In this case, the $\tdgd$ is
dyadic, the optimal profile is uniquely determined, and it and has no
gaps (the optimal profile is that of $G_1\concat G_1$).

\noindent\emph{Case $q<\half\,$.} Assume that $s\geq 2^k-2$, and that
there is a gap of size $g$ between signatures $s$ at level $d$, and
signatures $s+1$ at level $d+g+1$. Signatures $s+1$ may also be found
at level $d+g+2$. Without loss of generality, and by our assumption on
$s$, we can assume that there is a subtree of $\optq$ rooted at a node
$v$ at level $d+g+1-k$, and containing at least $2^k$ leaves with
signature $s+1$, including some at level $d+g+1$. Thus, we have
\begin{equation*} %\label{wt2}
\wt(v) \geq 2^k q^{s+1} > q^s = \wt(s),
\end{equation*}
the second inequality following from the definition of $k$. Therefore,
we must have $d + g+1-k \leq d$, or equivalently, $g \leq k-1$, for
otherwise exchanging $v$ and $s$ would decrease the cost,
contradicting the optimality of $\optq$.
\end{IEEEproof}

Next, we bound the rate of change of signature magnitudes as a function
of depth in an optimal tree. Together with the bound on gap sizes in
Lemma~\ref{lem:gaps}, this will lead to the proof of
Theorem~\ref{th:singular}. It follows from Lemma~\ref{lem:2levels} that
for every signature $s\geq 0$ there is a level of $\optq$ containing at
least one half of the $s+1$ leaves with signature $s$. We denote the
depth of this level by $L(s)$ (with some fixed policy for ties),
dependence on $\optq$ being understood from the context.

\begin{figure}
\centering{\includegraphics[width=3.4in]{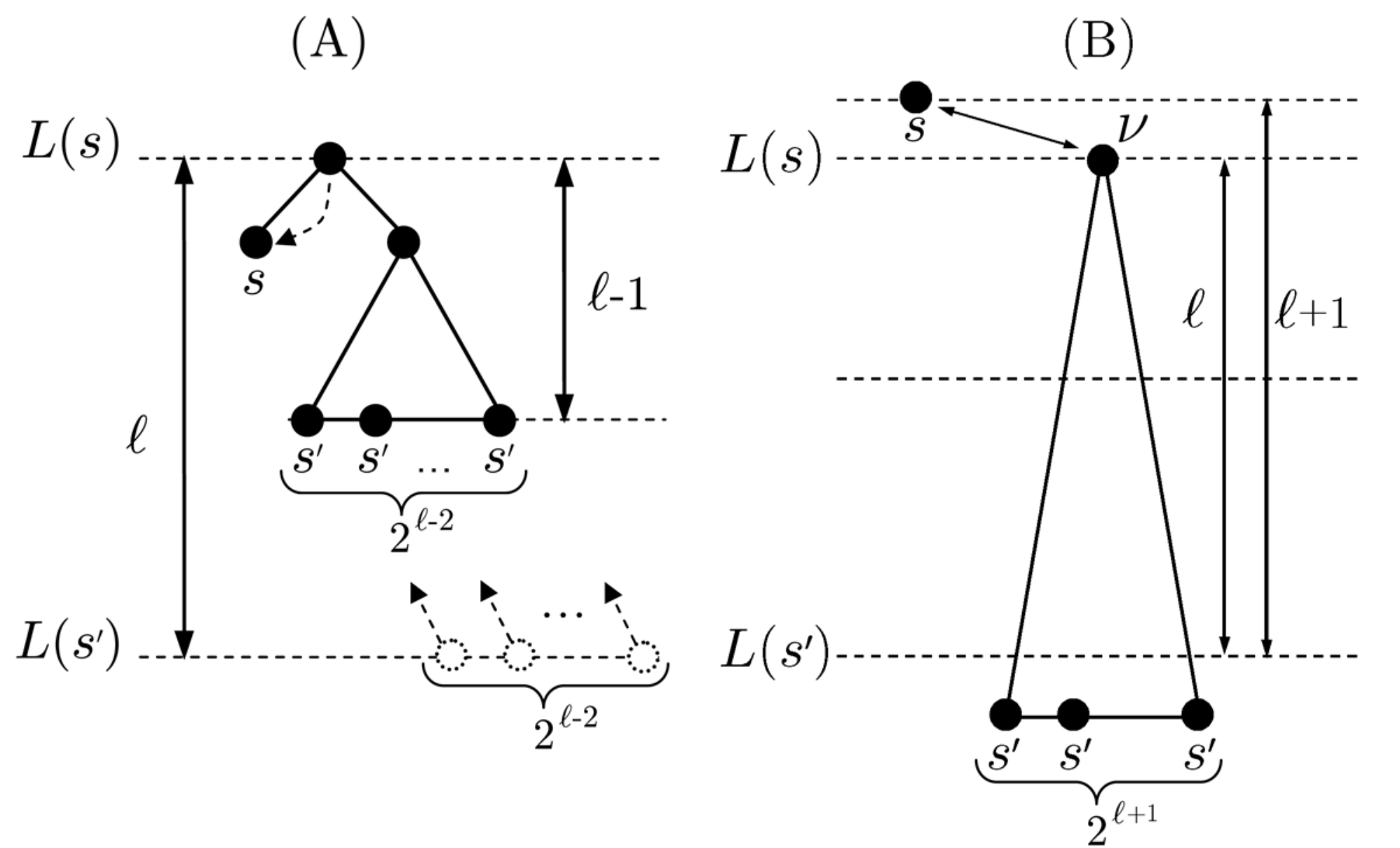}}
\caption{Tree transformations.\label{fig:transf}}
\end{figure}
\begin{lemma}\label{lem:bounds}
Let $s$ be a signature, and $\ell \geq 2$ a positive integer such that
$s\geq 2^{\ell+2}-1$, and such that $L(s') = L(s) + \ell$ for some
signature $s' > s$.  Then, for $\optq$, we have
\begin{equation}\label{eq:bounds}
\frac{\ell-2}{\log\qinv} \leq\; s'-s\; \leq
\frac{\ell+1}{\log\qinv}\;.
\end{equation}
\end{lemma}
\begin{IEEEproof}
Since $s'>s\geq 2^{\ell+2}-1 >  2^{\ell-1}-1$, by the definition of
$L(s')$, there are more than $2^{\ell-2}$ leaves with signature $s'$ at
level $L(s')$. We perform the following transformation (depicted in
Figure~\ref{fig:transf}(A)) on the tree $\optq$, yielding a modified
tree $\optq'$: Choose a leaf with signature $s$ at level $L(s)$, and
graft to it a tree with a left subtree consisting of a leaf with
signature $s$ (``moved'' from the root of the subtree), and a right
subtree that is a balanced tree of height $\ell-2$ with $2^{\ell-2}$
leaves of signature $s'$. These signatures come from $2^{\ell-2}$
leaves at level $L(s')$ of $\optq$, which are removed. It is easy to
verify that the modified tree $\optq'$ defines a valid, albeit
incomplete, code for the alphabet of a TDGD. Next, we estimate the
change, $\Delta$, in cost due to this transformation. We have
$$
\Delta = \cost{\optq'}-\cost{\optq} = q^s -
2^{\ell-2}q^{s'}\,.
$$
The term  $q^s$ is due to the increase, by one, in the code length for
the signature $s$, which causes an increase in cost, while the term
$-2^{\ell-2}q^{s'}$ is due to the decrease in code length for
$2^{\ell-2}$ signatures $s'$, which produces a decrease in cost.
Since $\optq$ is optimal, we must have $\Delta \geq 0$, namely,
$$
0 \leq q^s - 2^{\ell-2}q^{s'} =
q^s\left(1-2^{\ell-2}q^{s'-s}\right),
$$
and thus, $2^{\ell-2}q^{s'-s}\le 1$, from which the lower bound
in~\eq{eq:bounds} follows. (Note: clearly, the condition $s\geq
2^{\ell-1}-1$ would have sufficed to prove the lower bound; the
stricter condition of the lemma will be required for the upper bound,
and was adopted here for uniformity.)

To prove the upper bound, we apply a different modification to $\optq$.
Here, we locate $2^{\ell+1}$ signatures $s'$ at level $L(s')$, and
assume, without loss of generality, that these signatures are the
leaves of a balanced tree of height $\ell+1$, rooted at a node
$\subroot$ of depth $L(s)-1$. The availability of the required number
of leaves at level $L(s')$ is guaranteed by the conditions of the
lemma. We then exchange $\subroot$ with a leaf of signature $s$ at
level $L(s)$. The situation,
\emph{after} the transformation, is depicted in
Figure~\ref{fig:transf}(B). The resulting change in cost is computed as
follows.
$$
\Delta = \cost{\optq'}-\cost{\optq} = -q^s+
2^{\ell+1}q^{s'}\,.
$$
As before, we must have $\Delta \geq 0$, from which the upper bound
follows.
\end{IEEEproof}

We are now ready to prove Theorem~\ref{th:singular}.

\begin{IEEEproof}[Proof of Theorem~\ref{th:singular}]
We assume, without loss of generality, that $\qeps>q$, and we write
$\qeps = q(1+\eps)$, $0 < \eps < q^{-1}-1$. In $\optq$, choose a
sufficiently large signature $s$ (the meaning of ``sufficiently
large''
will be specified in the sequel), and a node of signature $s$ at level
$L(s)$. Let $s'>s$ be a signature such that $\ell \defined L(s')-L(s)
\ge 2$. We apply the transformation of Figure~\ref{fig:transf}(A) to
$\optq$, yielding a modified tree $\optq'$. We claim that when weights
are taken with respect to $\tdgd(\qeps)$, and with an appropriate
choice of the parameter $\ell$, $\optq'$ will have strictly lower cost
than $\optq$. Therefore, $\optq$ is not optimal for
$\tdgd(\qeps)$. To prove the claim, we compare the costs of
$\optq$ and $\optq'$ with respect to $\tdgd(\qeps)$. Reasoning as in
the proof of the lower bound in Lemma~\ref{lem:bounds}, we write
\begin{align}
\Delta &= \cost[\qeps]{\optq'}-\cost[\qeps]{\optq}= \qeps^s -
2^{\ell-2}\qeps^{s'} \nonumber\\
&=\qeps^s\left(1-2^{\ell-2}\qeps^{s'-s}\right) \leq
 \qeps^s\left(1-2^{\ell-2}\qeps^{\frac{\ell+1}{\log
 \qinv}}\right)\label{eq:chain}
\end{align}
where the last inequality follows from the upper bound in
Lemma~\ref{lem:bounds}. It follows from~\eq{eq:chain} that we can make
$\Delta$ negative if
$$
\ell-2 + \frac{\ell+1}{\log\qinv}\log\qeps > 0.
$$
Writing $\qeps$ in terms of $q$ and $\eps$, and after some algebraic
manipulations, the above condition is equivalent to
\begin{equation}\label{eq:ell}
\ell > 3\frac{\log\qinv}{\log(1+\eps)}-1\,.
\end{equation}
Hence, choosing a large enough value of $\ell$, we get $\Delta <0$,
and
we conclude that the tree $\optq$ is not optimal for $\tdgd(\qeps)$,
subject to an appropriate choice of $s$, which we discuss next.

The argument above relies strongly on Lemma~\ref{lem:bounds}. We recall
that in order for this lemma to hold, $\ell$ and the signature $s$ must
satisfy the condition $s\geq 2^{\ell+2}-1$. Now, it could happen that,
after choosing $\ell$ according to~\eq{eq:ell} and then $s$ according
to the condition of Lemma~\ref{lem:bounds}, the level $L(s)+\ell\,$
does not contain $2^{\ell-2}$ signatures $s'$ as required (e.g., when
the level is part of a gap). This would force us to increase $\ell$,
which could then make $s$ violate the condition of the lemma. We would
then need to increase $s$, and re-check $\ell$, in a potentially
vicious circle. The bound on gap sizes of Lemma~\ref{lem:gaps} allows
us to avoid this trap. The bound in the lemma depends only on $q$ and
thus, for a given TDGD, it is a constant, say $g_q$. Thus, first, we
choose a value  $\ell_0$ satisfying the constraint on $\ell$
in~\eq{eq:ell}. Then, we choose $s\geq 2^{\ell_0+g_q+4}$. Now, we try
$\ell=\ell_0,\ell_0+1, \ell_0+2,
\ldots,$ in succession, and check whether level $L(s)+\ell$ contains
enough of the required signatures. By Lemmas~\ref{lem:2levels}
and~\ref{lem:gaps}, an appropriate level $L(s')$ will be found for some
$\ell
\le\ell_0+ g_q+2$.  For such a value of $\ell$, we have
$2^{\ell+2}-1 \le 2^{\ell_0+g_q+4}-1 < s$, satisfying the condition of
Lemma~\ref{lem:bounds}. This condition, in turn, guarantees also that
there are at least $2^{\ell-2}$ signatures $s'$ at $L(s')$, as
required.
\end{IEEEproof}

\section{Optimal codes for  TDGDs with $q=\qparII$}
\label{secconst}\label{sec:q>=1/2}

It follows from the results of Section~\ref{sec:singular} that it is
infeasible to provide a compact description of optimal codes for
\tdgd{s} covering all values of the parameter $q$, as can be done with
one-dimensional geometric distributions~\cite{gol66,gv75} or their
two-sided variants~\cite{msw00}. Instead, we describe optimal prefix
codes for a discrete sequence of values of $q$, which provide good
coverage of the parameter range. %
In this section, we study optimal codes for TDGDs with parameters
$q=\qparII$ for integers $k\geq 1$, i.e., $q\ge\half$, while in
Section~\ref{sec:q<1/2} we consider parameters of the form $q=\qparI$,
$k>1$, covering the range $q<\half$ (the two parameter sequences
coincide at $k=1$, $q=\half$, which we choose to assign to the case
covered in this section).

\subsection{Initial characterization of optimal codes for $q=\qparII$
}
\label{sec:general_characterization}
The following theorem characterizes optimal codes for \tdgds\ of
parameter $q=\qparII$, $k \geq 1$, in terms of unary codes and Huffman
codes for certain finite distributions. In
Subsection~\ref{sec:toptree}
we further refine the characterization by providing explicit
descriptions of these Huffman codes.

\begin{theorem}
\label{theo:opt-code}
An optimal prefix code $C_k$ for $\tdgd(q)$, with $q=\qparII$, $k\geq
1$, is given by
\begin{equation*}
C_k(i, j) = T_k(i \bmod k, j \bmod k) \cdot \textstyle
G_1(\left\lfloor\frac{i}{k}\right\rfloor)\cdot
G_1(\left\lfloor\frac{j}{k}\right\rfloor ),
\end{equation*}
%\]
where $G_1$ is the unary code, and $T_k$, referred to as the \emph{top
code}, is an optimal code for the finite source defined by the
following symbol set and respective weights:
\begin{equation}\label{eq:topsource}
\Ak=\{ (i, j) \ | \ 0 \le i, j < k\},\quad w(i,j) = q^{i+j}\,.
\end{equation}
\end{theorem}

\textbf{Remarks.}
\begin{enumerate}
\item  Theorem~\ref{theo:opt-code} can readily be generalized to blocks
  of $d>2$ symbols. For simplicity, we present the proof for $d=2$.
\item\label{rem:2afterThm2} Notice that $C_k(i,j)$ concatenates
the ``unary'' parts of the codewords for $i$ and $j$ in a Golomb
  code of order $k$ (as if encoding $i$ and $j$ separately), but
  encodes the ``binary'' part jointly by means of $T_k$, which, in
  general, does not yield the concatenation of the respective
  ``binary'' parts $Q_k(i)$ and $Q_k(j)$. However, when $k=1$ and
  $k=2$, $C_k$ \emph{is} equivalent to the full concatenation
  $G_k\concat G_k$. When $k=1$, the code $T_k$ is void, and $C_1 =
  G_1\concat G_1$. The parameter in this case is $q=\half$, the
  geometric distribution is dyadic, and the code redundancy is
  zero. When $k=2$, we have $q=1/\sqrt{2}$ and the finite source
  $\Ak$ has four symbols with respective weights
  $\{\,1,\,\sqrt{2}/2,\,\sqrt{2}/2,\,1/2\,\}$. This source is
  quasi-uniform, and, therefore, it admits $Q_4$ as an optimal
  tree. This is a balanced tree of depth two, which can also be
  written as $Q_4 = Q_2 \concat Q_2$. Thus, we have
$\Cplus[2]=G_2\concat G_2$. Later on in the section, in
Corollary~\ref{cor:golomb-not-optimal}, we will show that this
situation will not repeat for larger values of $k$: the ``symbol by
symbol'' code $G_k\concat G_k$ is strictly suboptimal for
$\tdgd(2^{-1/k})$ when $k>2$.
\end{enumerate}

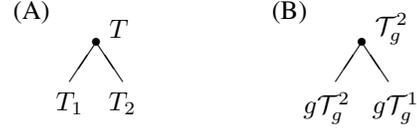
\begin{figure}
\begin{center}
\setlength{\unitlength}{0.05in}
\begin{picture}(40,15)(20,0)
\put(18.5,10){\makebox(0,0){(A)}}
\put(20,0){
{\hspace{-1em}\btree[$T$]{$T_1$}{$T_2$}}
}
\put(45.5,10){\makebox(0,0){(B)}}
\put(47,0){
$\btree[$\mcT^2_g$]{$\!\!\!\!g \mcT^2_g$}{$\;\;g\mcT^1_g$}$
}
\end{picture}
\end{center}
\caption{\label{fig:Tgram}Graphical representations for trees with associated weights.}
\end{figure}

\begin{figure}[t]
\begin{center}
\includegraphics[width=0.95\linewidth]{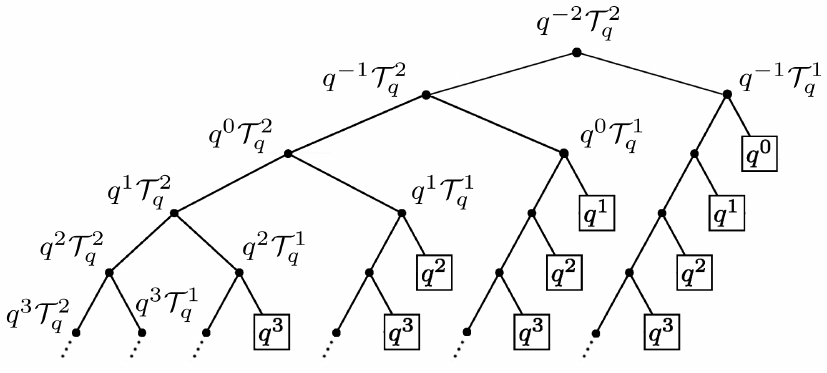}
\end{center}
\caption{\label{fig:T2}
The tree $q^{-2} \mathcal{T}^2_q$.}
\end{figure}

In deriving the proof of Theorem~\ref{theo:opt-code} and in subsequent
sections, we shall make use of the following notations to describe and
operate on some infinite trees with weights associated to their leaves.
We denote by $\boxed{\rule{0pt}{1.5ex}v}$ the trivial tree consisting
of a single node (leaf) of weight $v$. Given a tree $T$ and a scalar
$g$, $g T$ denotes the tree $T$ with all its weights multiplied by $g$.
Given trees $T_1$ and $T_2$, the graphic notation in
Figure~\ref{fig:Tgram}(A) represents a tree $T$ consisting of a root
node with $T_1$ as its left subtree and $T_2$ as its right subtree,
each contributing its respective leaf weights. The multiset of weights
associated with $T$ is the union of the multisets associated with $T_1$
and $T_2$. We will also use the notation $[\,T_1\;\; T_2\,]$ to
represent the forest consisting of the separate trees $T_1$ and $T_2$,
which has the same associated multiset of weights as the tree $T$ of
Figure~\ref{fig:Tgram}(A), but a different underlying graph. We denote
by $\mcT^1_g$ the tree of a unary code whose leaf at each depth $i\ge
1$ has weight $g^i$, and by $\mcT^2_g$ the structure in
Figure~\ref{fig:Tgram}(B).
It is readily verified that $\mcT^2_g$ corresponds to the concatenation
of two unary codes, with each of the $i-1$  leaves at depth $i \ge 2$
of $\mcT^2_g$ carrying weight $g^{i}$. In particular, as shown in
Figure~\ref{fig:T2}, the tree $q^{-2}\mathcal{T}^2_{q}$ corresponds to
the optimal tree for the dyadic \tdgd\ with $q=\half$, where each leaf
is weighted according to the signature of the symbol it encodes.

The following lemma follows directly from the above definitions,
applying elementary symbolic manipulations on geometric sums.
\begin{lemma}\label{lem:weights}
For any real number $g$, $0<g<1$, we have
 $\displaystyle w(\Tg{2}) = w(\Tg{1})^2
 = \left(\,\frac{g}{1-g}\,\right)^2$. In particular, if $q =
 \qparII$, we have $w(\mcT^2_{q^k}) = w(\mcT^1_{q^k}) = 1$.
 \end{lemma}

We rely on this observation in the proof of Theorem~\ref{theo:opt-code}
below. In the proof, when defining virtual symbols, we further overload
notation and regard trees with associated weights, such as
$q^r\mcT_{q^k}^d$, also as multisets of \emph{signatures}, with a
signature $s$ for each leaf of the tree with weight $q^s$.

\begin{IEEEproof}[Proof of Theorem~\ref{theo:opt-code}]
We use the Gallager-Van Voorhis construction~\cite{gv75}.  For $s\geq
0$, define the reduced source
\[
\mathcal{W}_s = \mathcal{H}_{s} \cup \mathcal{F}_s
\]
where
\[
\mathcal{H}_s= \{ i \in \alpb \ | \ i <
s\}
\]
(signatures in $\mathcal{H}_s$ occur with the same multiplicity as in
$\alpb$), and
\[ \mathcal{F}_s=\bigcup_{i=0}^{k-1} \{
\underbrace{q^{s+i}\mathcal{T}^2_{q^k}}_{\text{ $k$ times}},
\underbrace{q^{s+i}\mathcal{T}^1_{q^k}}_{\substack{s+k+i+1\\\text{times}}},
\, \underbrace{s+i}_{\substack{s+i+1\\ \text{times}}} \, \}.
\]
The multisets (of signatures) $q^{s+i}\mcT_{q^k}^1$ and $q^{s+i}\mcT_{q^k}^2$
play the role of virtual symbols in the reduced sources, as discussed in
Subsection~\ref{sec:gv} (we omit the qualifier `virtual' in the sequel). It
is readily verified that all the weights of symbols in $\mathcal{F}_s$ are
smaller than the weights of signatures in $\mathcal{H}_s$. Since $q=\qparII$,
by Lemma~\ref{lem:weights}, we have $\weight{q^{s+i}\mathcal{T}^2_{q^k}}=
\weight{q^{s+i}\mathcal{T}^1_{q^k}} = \weight{s+i}$. Thus, we can apply steps
of the Huffman procedure to $\mathcal{F}_s$ in such way that the $s+i+1$
signatures $s+i$ are merged with $s+i+1$ symbols
$q^{s+i}\mathcal{T}^1_{q^k}$, resulting in $s+i+1$ trees
$q^{s+i-k}\mathcal{T}^1_{q^k}$. The remaining $k$ symbols
$q^{s+i}\mathcal{T}^1_{q^k}$ can be merged with the $k$  symbols
$q^{s+i}\mathcal{T}^2_{q^k}$, resulting in $k$ trees
$q^{s+i-k}\mathcal{T}^2_{q^k}$ when $i$ ranges from $k{-}1$ down to $0$.
After this sequence of Huffman mergers,  $\mathcal{W}_s$ is transformed into
$\mathcal{W}_{s-k}$, as long as $s \geq k$. Starting from $s = t k$ for some
$t>0$, the procedure eventually leads to $\mathcal{W}_0$. Formally, our reduced source $\mathcal{W}_{tk},\;t \ge 0$, corresponds to
$\mathcal{S}_{t}$ in our description of the Gallager-Van Voorhis
construction in Section~\ref{sec:gv}. Thus, the iteration leads to
$\mathcal{S}_{0}$, as called for in the construction. It is readily verified that this source admits an additional sequence of Huffman mergers, as described above, leading (with a slight abuse of notation) to
\[
\mathcal{S}_{-1}=\bigcup_{i=0}^{k-1} \{
\underbrace{q^{i-k}\mathcal{T}^2_{q^k}}_{\substack{k\\\text{times}}},
\underbrace{q^{i-k}\mathcal{T}^1_{q^k}}_{\substack{i+1\\\text{times}}}
\,
\}\,.
\]
Continuing with the Huffman procedure, each symbol
$q^{i-k}\mathcal{T}_{q^k}^1$ in $\mathcal{S}_{-1}$ can be merged with a symbol
$q^{i-k}\mathcal{T}_{q^k}^2$, further leading, by the definition of $\Tg{2}$ (see
Figure~\ref{fig:Tgram}(B)), to a reduced source
\begin{align*}
\mathcal{S}^\ast=
\Bigl\{\;&
\underbrace{q^{-2k}\mathcal{T}^2_{q^k}}_{\substack{1\\ \text{time}}},\,
\underbrace{q^{-2k+1}\mathcal{T}^2_{q^k}}_{\substack{2\\
\text{times}}},\,
\underbrace{q^{-2k+2}\mathcal{T}^2_{q^k}}_{\substack{3\\
\text{times}}},
\dotsc\\
&\dotsc,\,\underbrace{q^{-k-1}\mathcal{T}^2_{q^k}}_{\substack{k\\
\text{times}}},
\underbrace{q^{-k}\mathcal{T}^2_{q^k}}_{\substack{k-1\\
\text{times}}},
\dotsc,\underbrace{q^{-3}\mathcal{T}^2_{q^k}}_{\substack{2\\ \text{times}}},
\underbrace{q^{-2}\mathcal{T}^2_{q^k}}_{\substack{1\\ \text{time}}} %
\;\Bigr\}\,.
\end{align*}
We now take a common ``factor'' $q^{-2k}\mathcal{T}_{q^k}^2$ from each
symbol~of $\mathcal{S}^\ast$. By the discussion of Figures~\ref{fig:Tgram}
and~\ref{fig:T2}, this factor corresponds to a copy of $G_1\concat
G_1$, with weights that get multiplied by $q^k$ every time the depth
increases by $1$. After the common factor is taken out, the source
$\mathcal{S}^\ast$ becomes the source $\Ak$ of~(\ref{eq:topsource}), to
which the Huffman procedure needs to be applied to complete the code
construction. Thus, the code described in the theorem is optimal.
\end{IEEEproof}

To make the result of Theorem~\ref{theo:opt-code} completely explicit,
it remains to characterize an optimal prefix code for the finite source
$\Ak$ of~(\ref{eq:topsource}). The following lemma presents some basic
properties of $\Ak$ and its optimal trees. Recall the definitions of
$\alpha$-uniformity and fringe thickness from Section~\ref{sec:prelim}.

\begin{lemma}\label{lem:4-uniform}
The source $\Ak$ is $4$-uniform, and it has an optimal tree $T$ of
fringe thickness $f_T\leq 2$.
\end{lemma}
\begin{IEEEproof} It follows from~(\ref{eq:topsource}) and the relation $q^k = \half$
that the maximal ratio between weights of symbols in $\Ak$ is
$q^{-2k+2} = 4q^2 < 4$. Hence, $\Ak$ is $4$-uniform. The claim on the
optimal tree holds trivially for $k\le 2$, in which case the optimal
tree for $\Ak$ is uniform. To prove the claim for $k>2$, consider the
multiset $\Akprime\subseteq \Ak$ consisting of the lightest
$2\lceil\frac{k(k-1)}{4}\rceil$ signatures in $\Ak$, i.e.,
\begin{align*}
\Akprime = \Kset \, \cup \,\bigl\{\,\underbrace{k,k,\ldots,k}_{k-1\text{ times}}\,,
&\underbrace{k{+}1,\ldots,k{+}1}_{k-2\text{ times}}\,,\dotsc\bigr.\\
&\bigl.\quad\dotsc,\, \underbrace{2k{-}3,2k{-}3}_{2 \text{ times}}\,,\,
\underbrace{2k{-}2}_{1 \text{ time}}\,
\bigr\}\,,
\end{align*}
where $\Kset=\{ k{-}1 \}$ if $k \bmod 4 \in\{2,3\}$, or $\Kset$ is empty
otherwise. The sum of the two smallest weights of signatures in $\Akprime$
satisfies
\begin{align*}
w(2k{-}2)+w(2k{-}3) & = q^{2k-2}+q^{2k-3}
= q^{2k-2}(1+q^{-1})\\
&  = \half(1+q^{-1})q^{k-2} > w(k-2)\,.
\end{align*}
The sum of the two largest weights in $\Akprime$, on the other hand, is
either $q^0$ if $k \bmod 4 \in \{ 0,1\}$, or $\half(1+q^{-1})$
otherwise. Therefore, if the Huffman procedure is applied to $\Ak$,
every pair of consecutive elements of $\Akprime$ will be merged,
without involving a previously merged pair. The ratio of the largest to
the smallest weight remaining after these mergers is at most
$\half(1{+}q^{-1})/q^{k{-}1} = q{+}1 < 2$. Hence, the resulting source
is quasi-uniform and has a quasi-uniform optimal tree. Therefore,
completing the Huffman procedure for $\Ak$ results in an optimal tree
of fringe thickness at most two.
\end{IEEEproof}

To complete the explicit description of an optimal tree for $\Ak$, we
will rely on a characterization of trees $T$ with $\fT\le 2$ that
are optimal for 4-uniform sources.\footnote{%
Notice that not every 4-uniform source admits an optimal tree with
$\fT\le 2$ (although the ones of interest in this section do). For
example, an optimal tree for the 4-uniform source with probabilities
$\frac{1}{10}(4,3,1,1,1)$ must have $\fT>2$.} This characterization is
presented next.
%%%%%%%%%%%%%%%%%%%%%%%%%%%%%%%%%%%%%%%%%%%%%%%%%%%%%%%%%%%%%%%%%%%%%%%%%%%%%%%%%%%%

\subsection{Optimal trees with $\fT\le 2$ for $4$-uniform sources}
\label{sec:fringe2}
To proceed as directly as possible to the construction of an optimal
tree for $\Ak$, we defer all the proofs of results in this subsection
to Appendix~\ref{app:fringe}. We start by characterizing all the
possible profiles for a tree $T$ with $N$ leaves, and $\fringeT
\leq 2$. Let $T$ be such a tree, let $m =\lceil\log N\rceil$, and
denote by $n_{\ell}$ the number of leaves at depth $\ell$ in $T$.

\begin{lemma}\label{lem:claimA} The profile of $T$ satisfies
$n_\ell = 0 $ for $\ell < m{-}2$ and  $\ell > m{+}1$, and either
$n_{m-2} = 0$ or $n_{m+1} = 0$ (or both, when $\fringeT\leq 1$).
\end{lemma}

It follows from Lemma~\ref{lem:claimA} that $T$ is fully characterized
by the quadruple $\left(n_{m-2},n_{m-1},n_m,n_{m+1}\right)$, with
either $n_{m-2}=0$ or $n_{m+1}=0$. We say $T$ is
\emph{long} if $n_{m-2}=0$, and that $T$ is \emph{short} if
$n_{m+1}=0$.  Defining $M = m - \short$, where $\short=1$ if $T$ is
short, or $0$ if it is long, a tree with $f_T
\le 2$ can be characterized more compactly by a triple of nonnegative
integers $\nT =\left(n_{M-1}, n_M, n_{M+1}\right)$. We will also refer
to this triple as the \emph{(compact) profile} of $T$, with the
associated parameters $N, m$, and $\short$ understood from the context.
Notice that when $n_{m-2}=n_{m+1}=0$, $T$ is the quasi-uniform tree
$Q_N$, and (abusing the metaphor), it is considered both long and short
(i.e., it has representations with both $\sigma=0$ and $\sigma=1$).

\begin{lemma}\label{prop:profiles1} Let $T$ be a tree with $\fT\le 2$.
For $\short\in\cerouno$ and $M = m-\short$,
define
\begin{equation*}%\label{eq:cminmax}
\ccmin = (N - 2^M)\short\quad\text{ and
}\quad
\ccmax = \left\lfloor\frac{2N-2^M}{3}\right\rfloor\,.
\end{equation*}
Then, $T$ is equivalent to one of the trees $\Tsc$ defined by the
profiles
\begin{align}
\nT[\Tsc] & = (n_{M{-}1},\,n_M,\, n_{M{+}1})\nonumber\\
& = \Big(2^M{-}N{+}c,\, 2N{-}2^M{-}3c,\,2c\Big),\nonumber\\
&\quad\quad\quad\quad\quad\short\in\cerouno,\;\;
\ccmin\le c \le \ccmax\,.\label{eq:cprofile}
\end{align}
\end{lemma}

\noindent\textbf{Remarks.}
\begin{enumerate}
\item
Equation~(\ref{eq:cprofile}) characterizes all trees with $N$
leaves and $\fT\le 2$ in terms of the parameters $\short$ and $c$.
The parameter $c$ has different ranges depending on $\short$: we
have $N-2^{m-1} \le c \le\lfloor\frac{2N-2^{m-1}}{3}\rfloor$ when
$\short=1$, and $0\le c
\le\lfloor\frac{2N-2^{m}}{3}\rfloor$ when $\short=0$. The use of the
parametrized quantities $M,\ccmin$, and $\ccmax$ will allow us to
treat the two ranges in a unified way in most cases. Also, notice
that $\Tsc[1,\,{\ccmin[1]}]$ and $\Tsc[0,\,{\ccmin[0]}]$ represent
the same tree, corresponding, respectively, to interpretations of
the quasi-uniform tree $Q_N$ as short or long.
\item\label{rem:c}
The parameter $c$ represents the number of internal (non-leaf)
nodes at level $M$ of $T$. An increase of $c$ by one corresponds to
moving a pair of sibling leaves previously rooted  at level $M-1$
to a new parent at level $M$ (thereby increasing the number of
internal nodes at that level by one). The number of leaves at level
$M$ decreases by three, and the numbers of leaves at levels $M-1$
and $M+1$ increase by one and two, respectively.
\end{enumerate}

Consider now a distribution on $N$ symbols, with associated vector of
probabilities (or weights) $\bldp=(\,p_1,p_2,\ldots,p_N \,)$, $p_1\geq
p_2\geq\cdots\geq p_N$. Let $\Lc$ denote the average code length of
$\Tc$ under $\bldp$ (with shorter codewords naturally assigned to
larger weights), and let
\begin{equation}\label{eq:defDc}
\DD = \Lc - \Lc[c-1],\quad \short\in\cerouno,\quad\ccmin < c \le \ccmax\,.
\end{equation}
It follows from these definitions, and the structure of the
profile~(\ref{eq:cprofile}) (see also Remark~\ref{rem:c} above), that
for $\short\in\cerouno$ and $\quad \ccmin  < c \le \ccmax$, we have
\begin{equation}
\label{eq:Dc}
\DD =  p_{N-2c+1} +p_{N-2c+2}- p_{2^M{-}N{+}c}\,.
\end{equation}
A useful interpretation of~(\ref{eq:Dc}) follows directly from the
profile~(\ref{eq:cprofile}): for $\Tc$, $\DD$ is the difference between
the sum of the two heaviest weights on level $M+1$ and the lightest
weight on level $M-1$.

Let $\sg(x)$ be defined as $-1,0$, or $1$, respectively, for negative,
zero, or positive values of $x$, and consider the following sequence
(recalling that $\ccmin[0]=0$):
\begin{align}
\sseq = {-}&\sg(\DDsc[{1,\ccmax[1]}]),\,{-}\sg(\DDsc[{1,\ccmax[1]-1}]),\,
\ldots,\,{-}\sg(\DDsc[1,{\ccmin[1]+1}]),\nonumber\\
&\sg(\DDsc[0,1]),\,\sg(\DDsc[0,2]),\,
\ldots,\,\sg(\DDsc[0,{\ccmax[0]}])\,.\label{eq:sseq}
\end{align}

\begin{lemma}\label{lem:cascade}
The sequence $\sseq$ is non-decreasing.
\end{lemma}

The definition of the sequence $\sseq$ induces a total ordering of the
pairs $(\short,c)$ (and, hence, also of the trees $\Tc$), with pairs
with $\short=1$ ordered by decreasing value of $c$, followed by pairs
with $\short=0$ in increasing order of $c$. The two subsequences
``meet'' at $\ccmin$, which defines the same tree regardless of the
value of $\short$ (in the pairs ordering, we take $(1,\ccmin[1])$ as
identical to $(0,\ccmin[0])=(0,0)$). We denote this total order by
$\preceq$. Recalling that the quantities $\DD$ are differences in
average code length between consecutive codes in this ordering,
Lemma~\ref{lem:cascade} tells us that, as we scan the codes in order,
we will generally see the average code length decrease monotonically,
reach a minimum, and then (possibly after staying at the minimum for
some number of trees) increase monotonically. In the following theorem,
we formalize this observation, and identify the trees $\Tc$ that are
optimal for $\bldp$.

\begin{table*}
\caption{\label{tab:choose}
Finding optimal trees $\Tc$ for $N=19$, %
$\bldp=\frac{1}{49}(4{,}4{,}3{,}3{,}3{,}3{,}3{,}3{,}3{,}3{,}3{,}2{,}2{,}2{,}2{,}
2{,}2{,}1{,}1)$ (optimal tree parameters emphasized in boldface).}
\begin{center}
\small
\renewcommand{\arraystretch}{1.25}
\begin{tabular}{|r|ccccccc|}
\hline
            &       &       &       &       &$\mathbf{(1,3)}=$&     &       \\[-1ex]
$(\short,c)$&$(1,7)$&$(1,6)$&$(1,5)$&$\mathbf{(1,4)}$&$\mathbf{(0,0)}\rule{1em}{0pt}$ &$\mathbf{(0,1)}$&$(0,2)$  \\
\hline
$(n_{M-1},n_M,n_{M+1})$&$(4,1,14)$&$(3,4,12)$&$(2,7,10)$&$\mathbf{(1,10,8)}$&$\mathbf{(13,6,0)}$&$\mathbf{(14,3,2)}$&$(15,0,4)$\\
\hline
$49\cdot\Lc$&$214$&$211$&$208$&$\mathbf{206}$&$\mathbf{206}$&$\mathbf{206}$&$208$ \\
\hline
$49\cdot\DD$&3&3&2&0&&0&2          \\
\hline
$\sseq$&-1&-1&-1&0&&0&1 \\[-1.4ex]
       &  &  &$(\shortm,\cmn)$  &$(\short_*,c_*)$ &&$(\short^*,c^*)$ &$(\shortp,\cpl)$\\
\hline
\end{tabular}
\end{center}
\end{table*}

\begin{theorem}\label{theo:choose}
Let $\bldp$ be a 4-uniform distribution such that $\bldp$ has an
optimal tree $T$ with $\fringeT\leq 2$. Define pairs $(\short_*,c_*)$
and $(\short^*,c^*)$ as follows:
\begin{eqnarray*}
(\short_*,c_*) &=& (1,\ccmax[1]) \quad \text{if } \DDsc[1,{\ccmax[1]}] \ge 0\,,\\
(\short^*,c^*) &=& (0,\ccmax[0]) \quad \text{if } \DDsc[0,{\ccmax[0]}]\le 0\,;
\end{eqnarray*}
otherwise, if $\DDsc[1,{\ccmax[1]}] < 0$, let $(\shortm,\cmn)$ be such
that $(-1)^{(\shortm)}\sg(\DDsc[{\shortm,\cmn}])$ is the last negative
entry in $\sseq$, and define
\begin{eqnarray*}
(\short_*,\,c_*) &=& (\shortm,\,\cmn -\,
\shortm)\,;\quad\quad
\end{eqnarray*}
if $\DDsc[0,{\ccmax[0]}] > 0$, let $(\shortp,\cpl)$ be such that
$(-1)^{(\shortp)}\sg(\DDsc[{\shortp,\cpl}])$ is the first positive
entry in $\sseq$, and define
\begin{eqnarray*}
(\short^*,\,c^*) &=& (\shortp,\,\cpl -1+\,\shortp)\,.
\end{eqnarray*}
Then, all trees $\Tc$ with
$(\short_*,c_*)\preceq (\short,c)
\preceq (\short^*,c^*)$ are optimal for $\bldp$.
\end{theorem}

Notice that, by Lemma~\ref{lem:cascade}, the range
$(\short_*,c_*)\preceq (\short,c) \preceq (\short^*,c^*)$ is well
defined and never empty, consistently with the assumptions of the
theorem and with Lemma~\ref{prop:profiles1}. The example in
Table~\ref{tab:choose} lists all the trees $\Tc$ with $\fT\le 2$ for
$N=19$, as characterized in Lemma~\ref{prop:profiles1}, and shows how
Theorem~\ref{theo:choose} is used to find optimal trees for a given
4-uniform distribution on $19$ symbols.

%%%%%%%%%%%%%%%%%%%%%%%%%%%%%%%%%%%%%%%%%%%%%%%%%%%%%%%%%%%%%%%%%%%%%%%%%%%%%%%%%%%%
\subsection{The top code}
\label{sec:toptree}
By Lemma~\ref{lem:4-uniform}, Theorem~\ref{theo:choose} applies to the
source $\Ak$ defined in~(\ref{eq:topsource}). We will apply the theorem
to identify parameters $\pairk$ that yield an optimal tree
$\Tsc[{\short_k,c_k}]$ for $\Ak$.

For the remainder of the section, we take $N=k^2$, and let
$\bldp=(p_1,p_2,\ldots,p_{k^2})$ denote the vector of (unnormalized)
symbol weights in $\Ak$, in non-increasing order. Thus, we have
$\bldp=(q^0,q^1,q^1,\ldots,q^j,q^j,\ldots,q^j,\ldots,q^{2k-3},q^{2k-3},q^{2k-2})$.
Here, $q^j$ is repeated $j+1$ times for $0\le j \le k{-}1$, and
$2k-1-j$ times for $k \le j \le 2k{-}2$. The following lemma, which
follows immediately from this structure, establishes the relation
between indices and weights in $\bldp$.

\begin{lemma}\label{lem:correspondence}
For $0 \leq  i <  k(k+1)/2$, we have $p_{i+1} = q^j$, where $j$ is the
unique integer in the range $0 \le j \le k-1$ satisfying
\begin{equation}\label{eq:large}
i = \frac{j(j+1)}{2}+r\quad\;\;\text{for some $r$},\;\;\;0 \le r \le j\,.
\end{equation}
For $0\le i' < k(k+1)/2$, we have $p_{k^2-i'} = q^{2k-2-j'}= \half
q^{k-2-j'}$, where $j'$ is the unique integer in the range $0\le j' \le
k-1$ satisfying
\begin{equation}\label{eq:small}
i'=\frac{j'(j'+1)}{2}+r'\quad\;\;\text{for some $r'$},\;\;\;0 \le r' \le j'\,.
\end{equation}
\end{lemma}

We define some auxiliary quantities that will be useful in the sequel.
Let $m=\lceil\log k^2\rceil$, $Q = k^2-\lceil {k(k-1)}/{4}\rceil$, and
$M' =\lceil\log_2 Q \rceil$, with dependence on $k$ understood from the
context. We assume that $k>2$, since the optimal codes for $k=1$ and
$k=2$ have already been described in
Subsection~\ref{sec:general_characterization}. It is readily verified
that we must have either $M'=m$ or $M'=m-1$. The next lemma shows that
the relation between $M'$ and $m$ determines the parameter $\short$ of
the optimal trees $\Tc$ for $\Ak$.
\begin{lemma}
\label{lem:dicho}
If $M'=m$, then trees $\Tc$ that are optimal for $\Ak$ are long
($\short=0$); otherwise, they are short ($\short=1$).
\end{lemma}
\begin{IEEEproof}
Assume $M'=m$. Then, we can write
\begin{align}
2^{m} & =  2^{M'} < 2^{1+\log \Qk} = 2\Qk\nonumber\\
& = 2k^2-2\lceil k(k-1)/4\rceil \le 2k^2-
k(k-1)/2\,,\label{eq:local1}
\end{align}
so $2^{m}-k^2 < k^2-k(k-1)/2$. If $\ccmin[1]+1 > \ccmax[1]$, then all
trees $\Tc$ in~(\ref{eq:cprofile}) are long. Otherwise,
$\DDsc[1,{\ccmin[1]{+}1}]$ is well defined, and we have
\begin{align}
-&\DDsc[1,{\ccmin[1]{+}1}] = -\DDsc[1,k^2-2^{m-1}+1]\nonumber\\
 & \quad =\, p_1 - (p_{2^m-k^2-1}+p_{2^m-2^k})\nonumber\\
 &\quad \le \,p_1 - 2 p_{k^2-k(k-1)/2} = p_1 -
2q^{k-1} = 1 - q^{-1} < 0\,,\label{eq:DDsc1}
\end{align}
where the first and second equalities follow from the definition of
$\ccmin[1]$ and from~(\ref{eq:Dc}), the first inequality from the
ordering of the weights and from~(\ref{eq:local1}), the third equality
from Lemma~\ref{lem:correspondence}, and the last equality from the
relation $q^k{=}\half$. By Lemma~\ref{lem:cascade}, we conclude that
optimal trees for $\Ak$ are long in this case. Similarly, when
$M'=m-1$, we have
\begin{equation}\label{eq:local2}
2^m \ge 2\Qk \ge 2k^2 -k(k-1)/2-2\,,
\end{equation}
so $2^m-k^2+1\ge k^2-k(k-1)/2-1$, and $p_{2^m-k^2+1} \le
p_{k^2-k(k-1)/2-1} =q^k =
\half$. If $\ccmax[0] = \ccmin[0] = 0$, then all trees $\Tc$ in~(\ref{eq:cprofile})
are short. Otherwise, similarly to~(\ref{eq:DDsc1}), we have
\[
\DDsc[0,1] = p_{k^2-1}+p_{k^2} - p_{2^m-k^2+1} > 2q^{2k-2} - \half =
\frac{q^{-2}}{2}-\half  > 0,
\]
which implies that optimal trees are short in this case.
\end{IEEEproof}
It follows from Lemma~\ref{lem:dicho} that we can take $m - M'$ as the
parameter $\short$ for all trees $\Tc$ that are optimal for $\bldp$.
Notice that $M'$ is analogous to the parameter $M$ defined in
Lemma~\ref{prop:profiles1}, but slightly stricter, in that, in cases
where a quasi-uniform tree is optimal, $m-M'$ will assume a definite
value in $\{0,1\}$ (which will vary with $k$), while, in principle, a
representation with either value of $\short$ is available. This very
slight loss of generality is of no consequence to our derivations, and,
in the sequel, we will identify $M$ with $M'$, i.e., we will take $M =
\lceil\log Q\rceil$. It also follows from Lemma~\ref{lem:dicho} that when applying
Theorem~\ref{theo:choose} to find optimal trees for $\bldp$, we only
need to focus on one of the two segments (corresponding to $\short{=}0$
or $\short{=}1$) that comprise the sequence $\sseq$ in~(\ref{eq:sseq}),
the choice being determined by the value of $k$. This will simplify the
application of the theorem.

Lemmas~\ref{lem:correspondence} and~\ref{lem:dicho}, together with
Theorem~\ref{theo:choose}, suggest a clear way, at least in principle,
for finding an optimal tree $\Tc$ for $\Ak$. The parameter $\short$ is
determined immediately as $\short = m - M$ (recalling that $m$ and $M$
are determined by $k$). Now, recalling the expression for $\DD$
in~(\ref{eq:Dc}), we observe that as $c$ increases, the weights
$p_{k^2-2c+1}$ and $p_{k^2-2c+2}$ also increase, while
$p_{2^M{-}k^2{+}c}$, which gets subtracted, decreases. Thus, since, by
Theorem~\ref{theo:choose}, an optimal value of $c$ occurs when $\DD$
changes sign, we need to search for the value of $c$ for which the
increasing sum of the first two terms ``crosses'' the value of the
decreasing third term. This can be done, at least roughly, by using
explicit weight values from Lemma~\ref{lem:correspondence} with
$i'\in\{2c-1,\,2c-2\}$ and $i=2^m-k^2+c$, and solving a quadratic
equation, say, for the parameter $j$ (the parameter $j'$ will be tied
to $j$ by the constraint $\DD\approx 0$). A finer adjustment of the
solution is achieved with the parameters $r$ and $r'$, observing that a
change of sign of $\DD$ can only occur near locations where the weights
in $\bldp$ change (i.e., ``jumps'' in either $j$ or $j'$), which occur
at intervals of length up to $k$. At the ``jump'' locations, either $r$
or $r'$ must be close to zero. While there is no conceptual difficulty
in these steps, the actual computations are somewhat involved, due to
various integer constraints and border cases. Theorem~\ref{theo:top}
below takes these complexities into account and characterizes,
explicitly in terms of $k$, the parameter pair $\pairk$ of an optimal
code $\Tskck$ for $\Ak$.

\begin{table}
\caption{\label{tab:code_examples}Optimal code parameters and profiles
for $\Ak,\;\,3\le k \le 10$.}
\begin{center}
\small
\renewcommand{\arraystretch}{1.0}
\begin{tabular}{|rrrrrrc|}
\hline
   $k$ &    $M$ &   $j$ &    $r$ & $\short_k$ &   $c_k$ &{\footnotesize $(n_{M-1},n_M,n_{M+1})$} \\
\hline
   2 &    2 &    0 &    0 & 0 &  0 & $(0, 4, 0)$\\
   3 &    3 &    0 &    0 & 1 &  1 & $(0, 7, 2)$\\
   4 &    4 &    1 &    0 & 0 &  1 & $(1, 13, 2)$\\
   5 &    5 &    3 &    1 & 0 &  0 & $(7, 18, 0)$\\
   6 &    5 &    1 &    0 & 1 &  5 & $(1, 25, 10)$\\
   7 &    6 &    5 &    0 & 0 &  0 & $(15, 34, 0)$\\
   8 &    6 &    2 &    2 & 0 &  5 & $(5, 49, 10)$\\
   9 &    6 &    0 &    0 & 1 & 17 & $(0, 47, 34)$\\
  10 &    7 &    7 &    1 & 0 &  1 & $(29, 69, 2)$\\
\hline
\end{tabular}
\end{center}
\vspace{-5mm}
\end{table}
\begin{theorem}\label{theo:top}\label{theo:topoftree}
Let $q= \qparII$, $\Qk = k^2-\lceil {k(k-1)}/{4}\rceil$, $m=\lceil\log
k^2\rceil$, and $M =\lceil\log Q \rceil$. Define the function
\begin{equation}\label{eq:deltax}\squeeze
\fxdelta(x) = 2k^2-2^{M+1} + x(x+1) - \frac{(k-x-2)(k-x-1)}{2}\,.
\end{equation}
 Let $x_0$ denote the largest real root of $\fxdelta(x)$, and let $\xi=\left
\lfloor x_0\right\rfloor$. Set
\begin{equation}\label{eq:jrk1}
(j,r) = \begin{cases}
\;\left(\,\xi,\,\left\lfloor
\frac{-\fxdelta(j)+1}{2}\right\rfloor
\,\right),&\text{if }\,\fxdelta(\xi) \le 2\xi ,\\
\;\rule{0pt}{2em}\bigl(\,\xi{+}1,\,0\,\bigr),& \text{otherwise}.
\end{cases}
\end{equation}
Then, the tree $\Tsc[{\short_k,\cck}]$, as defined by the
profile~(\ref{eq:cprofile}) with $\short = \short_k = m-M$ and
\begin{equation}\label{eq:c}
c = \cck = k^2-2^M+\frac{j(j+1)}{2} +r\,,
\end{equation}
is optimal for $\Ak$. Furthermore, $\cck$ is the smallest value of $c$
for any optimal tree $\Tsc[{\short_k,c}]$ for $\Ak$.
\end{theorem}

The  proof of Theorem~\ref{theo:top} is presented in
Appendix~\ref{app:proof_of_top}. In the theorem (and its proof), we
have chosen to identify the optimal tree $\Tsc[{\short_k,c}]$ with the
\emph{smallest} possible value of $c$. It can readily be verified that
this choice minimizes the \emph{variance} of the code length among all
optimal trees $\Tsc[{\short_k,c}]$. With only minor changes in the
construction and proof, one could also identify the \emph{largest}
value of $c$ for an optimal tree, and, thus, the full range of values
of $c$ yielding optimal trees $\Tsc[{\short_k,c}]$. For conciseness, we
have omitted this extension of the proof.

Examples of the application of Theorem~\ref{theo:top} are presented in
Table~\ref{tab:code_examples}, which lists the parameters $M$, $j$,
$r$, $\short_k$, $c_k$, and the profile of the optimal tree
$\Tsc[{\short_k,\,c_k}]$ defined by the theorem, for $3 \le k \le 10$.

The tools derived in the proof of Theorem~\ref{theo:top} also yield the
following result, a proof of which is also presented in
Appendix~\ref{app:proof_of_top}.

\begin{corollary}\label{cor:golomb-not-optimal}
Let $k>2$ and $q=2^{-1/k}$. Then, $G_k\concat G_k$ is not optimal for
$\tdgd(q)$.
\end{corollary}

\subsection{Average code length}\label{sec:averagecodelength}

The following corollary gives explicit formulas for the average code
length of the codes $\Cplus$ characterized in
Theorem~\ref{theo:opt-code} and Theorem~\ref{theo:top}. The proof is
deferred to Appendix~\ref{app:averagecodelength}.

\begin{corollary}\label{cor:averagelength}
Let $M$, $\fxdelta(x)$, $j$, and $r$ be as defined in
Theorem~\ref{theo:top}. Then, the average code length $\ncost{\Cplus}$
for the code $\Cplus$ under $\tdgd(q)$, for arbitrary $q$, is given by
\begin{equation}
\ncost{\Cplus}= M +  1 + \frac{{q}^{j} \FFF(q) }{(1-q^k)^2}\,,
\label{eq:avglenCk2}
\end{equation}
where
\begin{align*}
\FFF(q) & =
1-{q}^{k+1}+ ( 1-q )  \Big( {q}^{k+1}
 \left( k-j-1 \right) +j \Big)\\
 &  + ( 1-q ) ^{2} \Big( {q}^
{k} \big(\, 2\,r+\fxdelta( j) \, \big) -r \Big) \,.
\end{align*}
When $q=\qparII$, we have
\begin{equation}\label{eq:avglenCkqk}
 \ncost{\Cplus} =
M+1
 +2\,{q}^{j} \FFFP(q)\,,
\end{equation}
with
\[
\FFFP(q) =  1+ ( 1{-}q )  \bigl( q\,k+ ( 2{-}q
 ) j \bigr) + \left( 1-q \right) ^{2} \left( 1+\fxdelta \left( j
 \right)  \right)\,.
\]
\end{corollary}

\section{Optimal codes for TDGDs with
$q=\qparI$}\label{sec:firstconst}
\label{sec:q<1/2}

\subsection{The codes}
Assume $q=\qparI$ for some integer $k>1$. We reuse the notation
$\Ctree_m = Q_{2^m}$ for a uniform tree of depth $m$, assuming,
additionally, that its $2^m$ leaves have weight one. The infinite tree
(and associated multiset of leaf weights) $\Ltreek$ is recursively
defined as follows. Start from $\Ctree_k$, and attach to its leftmost
leaf a copy of $q\Ltreek$. Thus, $\Ltreek$ has $2^k{-}1$ leaves of
weight $q^s$ at depth $(s+1)k$ for all $s{\geq}0$, and no other leaves.
The related tree $\Ltreem$ is defined by starting from $\Ctree_{k-1}$,
and attaching to its leftmost leaf a copy of $q
\Ltreek$. Thus, $\Ltreem$ has $2^{k-1}{-}1$ leaves of weight $q^0$ at
depth $k-1$, and $2^k-1$ leaves of weight $q^s$ at depth $(s+1)k-1$
for
all $s>0$. The trees $\Ltreek$ and $\Ltreem$ are illustrated in
Figure~\ref{fig:ltrees}.

\begin{figure}[t]
\begin{center}
%\framebox%
{
\setlength{\unitlength}{1.45pt}
\begin{picture}(160,70)(28,-27.5)
\newsavebox{\ltreek}
\savebox{\ltreek}{
\put(-11,-11){\line(-1,-1){16}}
\put(-11,-11){\line(+1,-1){16}}
\put(-11,-11){\circle*{2}}
\put(-11,-21){\makebox(0,0){\footnotesize$\Ctree_k$}}
\put(-27,-27){\line(1,0){8}}
\put(-3,-27){\line(1,0){8}}
\qbezier[8](-19,-27)(-11,-27)(-3,-27)
\multiput(-27,-27)(4,0){3}{\circle*{2}}
\multiput(-3,-27)(4,0){3}{\circle*{2}}
\put(-23,-32.5){\makebox(0,0)[l]{\tiny$\underbrace{\hspace{28.5\unitlength}}_{2^k-1}$}}
\put(-35,-28.42){
    \put(0,2){\line(1,0){4}}
    \put(0,17.414){\line(1,0){4}}
    \put(2,6.5){\vector(0,-1){4.5}}
    \put(2,12.414){\vector(0,1){5}}
    \put(2,9.5){\makebox(0,0){\tiny$k$}}
}
}%
\newsavebox{\ltreem}
\savebox{\ltreem}{
\put(-11,-11){\line(-1,-1){12}}
\put(-11,-11){\line(+1,-1){12}}
\put(-11,-11){\circle*{2}}
\put(-10.5,-19.5){\makebox(0,0){\footnotesize$\Ctree_{k{-}1}$}}
\put(-23,-23){\line(1,0){6}}
\put(-5,-23){\line(1,0){6}}
\qbezier[8](-19,-23)(-11,-23)(-5,-23)
\multiput(-23,-23)(4,0){2}{\circle*{2}}
\multiput(-3,-23)(4,0){2}{\circle*{2}}
\put(-22.5,-28.5){\makebox(0,0)[l]{\tiny$\underbrace{\hspace{21\unitlength}}_{\;\;\;\quad\quad
2^{k{-}1}-1}$}}
\put(-35,-25){
    \put(0,2){\line(1,0){4}}
    \put(0,15){\line(1,0){4}}
    \put(2,6){\vector(0,-1){4}}
    \put(2,10){\vector(0,1){5}}
    \put(2,8.){\makebox(0,0){\tiny$k{-}1$}}
}
}%
\put(0,0){
  \put(88,38){\makebox(0,0){$\Ltreek$}}
  \multiput(100,40)(-16,-16){3} {
    \put(0,0){\usebox{\ltreek}}
  }
  \put(-2,0){
    \put(38,-22){\qbezier[8](0,0)(-5,-5)(-10,-10)}
    \put(110,13){\makebox(0,0)[l]{\footnotesize$q^0$}}
    \put(94,-3){\makebox(0,0)[l]{\footnotesize$q^1$}}
    \put(78,-19){\makebox(0,0)[l]{\footnotesize$q^2$}}
    \put(76,-24){\qbezier[8](0,0)(-5,-5)(-10,-10)}
  }%
}%
\put(80,0) {
  \put(85,36){\makebox(0,0){$\Ltreem$}}
  \put(96,36){\put(0,0){\usebox{\ltreem}}}
  \multiput(84,24)(-16,-16){2} {
    \put(0,0){\usebox{\ltreek}}
  } %
  \put(-2,0){
    \put(38,-22){\qbezier[8](0,0)(-5,-5)(-10,-10)}
    \put(105,13){\makebox(0,0)[l]{\footnotesize$q^0$}}
    \put(94,-3){\makebox(0,0)[l]{\footnotesize$q^1$}}
    \put(78,-19){\makebox(0,0)[l]{\footnotesize$q^2$}}
    \put(76,-24){\qbezier[8](0,0)(-5,-5)(-10,-10)}
  }%
}
\end{picture}
}
\end{center}
\caption{Trees $\Ltreek$ and $\Ltreem$.\label{fig:ltrees}}
\end{figure}

We describe a sequence of binary trees (and codes) $\Cminus$, which,
later in the section, will be shown to be optimal for TDGDs  with $q =
2^{-k}$, $k>1$. We describe the trees by \emph{layers}. A layer
$\layer$ is a collection of consecutive levels of the tree, containing
all the leaves with signature $s$.  The structure of the layers, and
how $\layer$ unfolds into $\layer[s+1]$ for all $s$, are presented
next, providing a full description of the trees $\Cminus$.

Assume $k > 1$ is fixed. We distinguish two main cases for the
structure of $\layer$, which depend on the value of $s$, as specified
below. In the description of the layers, each tree structure is a
virtual symbol. We will refer to both original and virtual symbols
simply as \emph{symbols}.

\medskip

\textbf{Case 1)} $\; 0 \le s \le 2^{k-1}-2$:

\noindent
Write $s = 2^i+j-1$ with $0\le i \le k-2,\; 0\le j
\le 2^i-1$. Layer $\layer$ consists of nodes in two levels,
arranged as follows:
\begin{equation}\label{eq:case1}
q^s{\cdot}\Bigg[ %
\,\leafrep{2^{i}-j-1}
\tree{\small$\rsymb$}{\small\boxed{1}}%
\treerep{\boxed{1}}{\boxed{1}}{j}
\Bigg] %
\end{equation}
(recall that the factor $q^s$ multiplies all the weights of objects inside the brackets, so that the leaves denoted $\boxed{1}$ in~(\ref{eq:case1}) indeed correspond to signatures $s$).

The symbol $\rsymb$ represents a tree containing all the signatures
strictly greater than $s$, scaled by $q^{-s}$. Layer $\layer$ emerges
from constructing a quasi-uniform tree for $s+2$ symbols ($s+1$
signatures $s$, and the symbol $\rsymb$), attached to $\rsymb[s-1]$ of
the previous layer if $s>0$, or to the root of the tree if $s=0$. We
have $s+2 = 2^i+1+j$, $0\le j\le 2^i-1$, so the quasi-uniform tree has
$2^i-j-1$ leaves at depth $i$, and $2j+2$ leaves at level $i+1$, as
shown in~(\ref{eq:case1}).

\medskip

\textbf{Case 2)} $\; s \ge 2^{k-1}-1$:

\noindent
Write
\begin{equation}\label{eq:s}
s= 2^{k-1}{-}1+(2^k-1)\ell +j,\;\;\ell\ge 0,\;\; 0 \le j
<2^{k}-1\,.
\end{equation}
There are five types of layers in this case, as described below. The
symbol $\rsymb$ in each case represents a tree containing all the
signatures strictly greater than $s$ that are not contained in other
virtual symbols in $\layer$, suitably scaled by $q^{-s}$. Also, it will
be convenient to use the notation $\macrosymbol$ as shorthand for the
sequence
\begin{equation}\label{eq:macro}
\macrosymbol :\quad
q \Ltreek\,,
\;\;\leafrep[-.5ex]{2^{k}{-}1} \;\;
\end{equation}
($\macrosymbol$ still counts as $2^k$ symbols in $\layer$).

\medskip

(i) $0\le j \le 2^{k-1}{-}3$ (for $k>2$):
\begin{equation}\label{eq:(i)}
  q^s{\cdot}\Bigg[\,
\stackT{\ell}
\leafrep{ 2^{k-1}-j-1}
\tree{$\!\!\rsymb$}{$\boxed{1}$}
\treerep{$\boxed{1}$}{$\boxed{1}$}{j}
\Bigg]
\end{equation}

\bigskip

\noindent(ii)
{$ j = 2^{k-1}{-}2:$}
\begin{equation}\label{eq:(ii)}
q^s\cdot\Bigg[\;
\stackT{\ell}\;\quad
\tree{$q\,\Ctree_{k-1}$}{$\quad\;\rsymb$}\quad
\treerep{$\boxed{1}$}{$\boxed{1}$}{2^{k-1}{-}1}
\;\Bigg]
\end{equation}

\bigskip

\noindent(iii) {$ 2^{k{-}1}-1 \le j \le 2^{k}{-}4$:}
\begin{equation}\label{eq:(iii)}
 q^s \cdot \Bigg[\;
\stackT{\ell}
\leafrep{3\cdot 2^{k{-}1}{-}2{-}j}
\tree{$q\,\Ctree_{k-1}$}{$\quad\;\rsymb$}
\;\treerep{$\boxed{1}$}{$\boxed{1}$}{j{-}2^{k-1}{+}1}
\Bigg]
\end{equation}

\bigskip

\noindent(iv) {$j = 2^{k}{-}3$:}
\begin{equation}\label{eq:(iv)}
q^s \cdot\Bigg[\,
\stackT{\ell}
\leafrep{2^{k-1}{+}1}
\tree{$q \Ltreem$}{$\quad\;\rsymb$}
\treerep{$\boxed{1}$}{$\boxed{1}$}{2^{k{-}1}{-}2}
\,\Bigg]
\end{equation}

\bigskip

\noindent(v) {$j = 2^{k}{-}2$:}
\begin{equation}\label{eq:(v)}
q^s{\cdot}\Bigg[\,
\stackT{\ell}
\;\; q\Ltreek %
\leafrep{2^{k{-}1}{-}1}
\!\!\tree{$\rsymb$}{$\boxed{1}$}
\!\!\treerep{$\boxed{1}$}{$\boxed{1}$}{2^{k{-}1}{-}1}
\,\Bigg]
\end{equation}
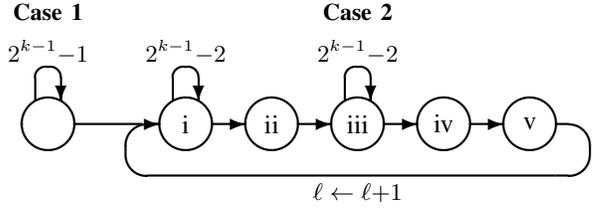
\begin{figure}
\begin{center}
\setlength{\unitlength}{0.009in}

\begin{picture}(320,120)(-20,05)
\thicklines
\put(0,40){%
\put(0,65){\makebox(0,0){\small\textbf{Case 1}}}%
\put(0,0){\circle{30}}%
\put(0,13.25){\oval(15,40)[t]} %
\put(7.5,18.25){\vector(0,-1){5}} %
\put(0,42){\makebox(0,0){\small$2^{k-1}{-}1$}}%
\put(15,0){\line(1,0){40}}
}%
\put(80,40) { %
\put(100,65){\makebox(0,0){\small\textbf{Case 2}}}%
\multiput(0,0)(50,0){5}{\circle{30}}%
\multiput(15,0)(50,0){4}{\vector(1,0){20}}%
\put(215,-15){\oval(40,30)[r]}%
 \put(215,-30){\line(-1,0){230}}%
 \put(100,-40){\makebox(0,0){\small$\ell\leftarrow \ell{+}1$}}
 \put(-15,-15){\oval(40,30)[l]}%
 \put(-20,0){\vector(1,0){5}}%
 \put(0,42){\makebox(0,0){\small$2^{k-1}{-}2$}}%
 \put(0,13.25){\oval(15,40)[t]}
 \put(7.5,18.25){\vector(0,-1){5}}
\put(100,0){ %
 \put(0,42){\makebox(0,0){\small$2^{k-1}{-}2$}}%
 \put(0,13.25){\oval(15,40)[t]}
 \put(7.5,18.25){\vector(0,-1){5}}
}
 \put(0,0){\makebox(0,0){i}}
 \put(50,0){\makebox(0,0){ii}}
 \put(100,0){\makebox(0,0){iii}}
 \put(150,0){\makebox(0,0){iv}}
 \put(200,0){\makebox(0,0){v}}
}%
\end{picture}
\end{center}
\caption{\label{fig:layertransitions}Layer transitions in $\Cminus$ for
$k>2$. The expressions above the self-loops indicate the number of iterations
on the given layer type before the transition to the next type.}
\end{figure}

%%%%%%%%%%%%%%%%%%%%%%%%%%%%%%%%%%%%%%%%%%%%%%%%%%%%%%%%%%%%%%%%%%%%%%%%%%%%%%%%%
The last layer from Case 1 contains all the signatures $s'=2^{k-1}-2$. All
signatures $s > s'$ are contained in $\rsymb[s']$. In particular, there are
$2^{k-1}$ signatures $s'+1=2^{k-1}-1$. Assume $k>2$. A quasi-uniform tree
with $2^{k-1}+1$ leaves is constructed, rooted at $\rsymb[s']$. This tree has
$2^{k-1}-1$ leaves labeled ${s'+1}$ at depth $k-1$ from its root, and two
leaves at depth $k$, one of which is labeled ${s'+1}$, and one that serves as
the root for $\rsymb[s'+1]$. This is consistent with the structure of the
first layer in Case 2 shown in~(\ref{eq:(i)}), with $s=s'+1$, $\ell=0$ and
$j=0$. From that layer on, layers of types (i)--(v) above unfold following
the cyclic pattern shown in Figure~\ref{fig:layertransitions}. Layers of
types (i) and (iii) are repeated $2^{k-1}{-}2$ times each in the cycle, which
is closed by a transition from a layer of type (v) back to one of type (i),
corresponding to an increment of the value of $\ell$ by one.

When $k=2$, layers of type (i) or (iii) are not used. In this case,
the only layer in Case 1 contains the signature $0$. A uniform tree
$\Ctree_2$ is constructed, rooted at $\rsymb[0]$. One pair of sibling
leaves is assigned to signature $1$, while the other pair is assigned
to $\rsymb[1]$ and $\Ctree_1$, attaining a configuration of type (ii)
in Case~2. From that point on, the cyclic layer sequence is
(ii)$\to$(iv)$\to$(v)$\to$(ii).

\begin{figure}[t]
\begin{center}
\includegraphics[width=3.45in]{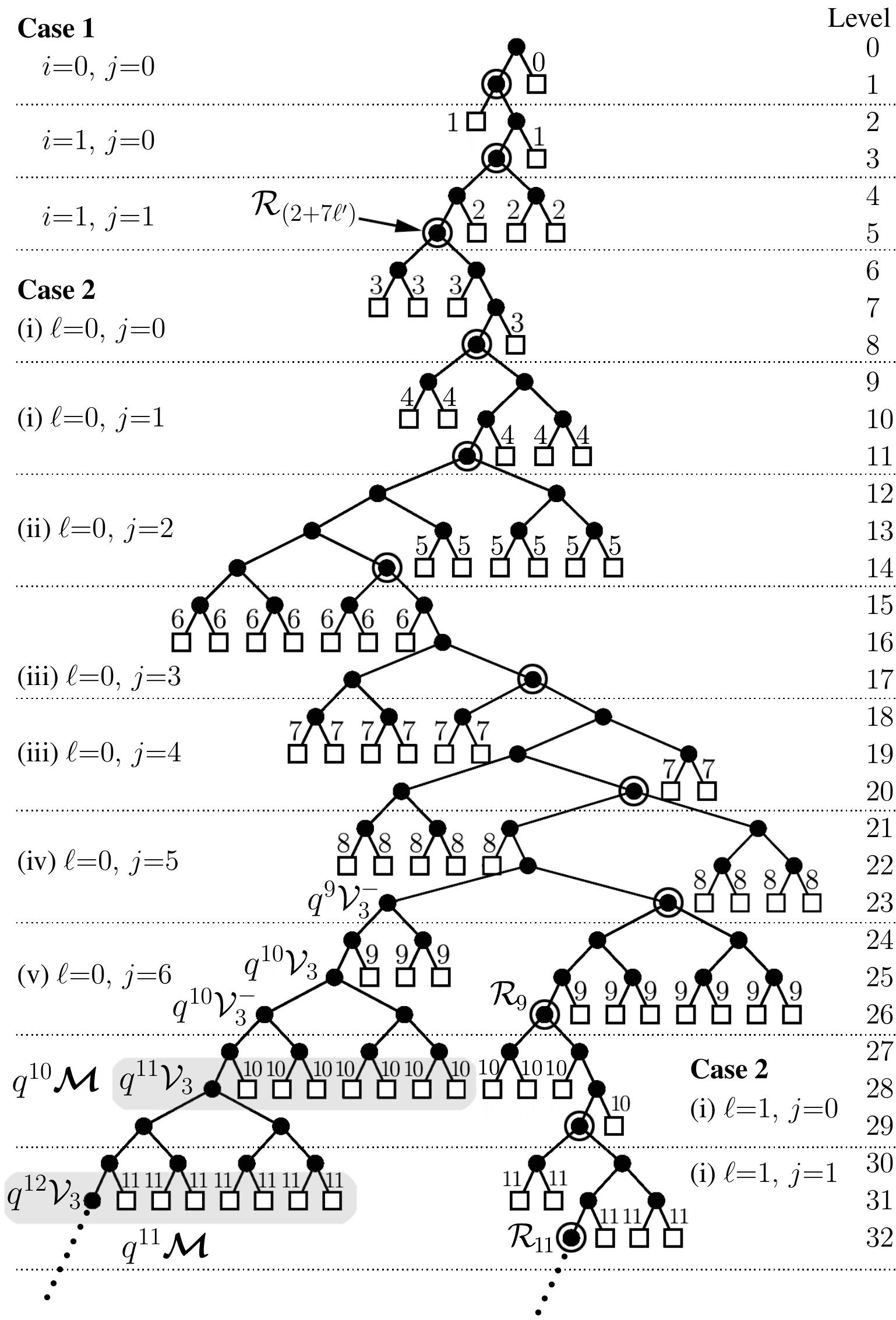}%
\end{center} %
\caption{\label{fig:code_1_8} Top levels comprising layers $\layer[s]$
for $s \le 11$ in the optimal tree
$\Cminus[3]$ ($q=\eighth$). Leaf signatures are noted;
dotted lines separate layers $\layer$,
and circled nodes represent roots of trees $\rsymb$. Grayed
ovals represent sequences $q^s\macrosymbol$.}
\end{figure}

The fine details of the various layer transitions, justifying the structure
in Figure~\ref{fig:layertransitions}, are given in Appendix~\ref{app:layers}.
The structure is also illustrated by the example in
Figure~\ref{fig:code_1_8}, which shows the layers $\layer$ for $s \le 11$ in
$\Cminus[3]$.

%%%%%%%%%%%%%%%%%%%%%%%%%%%%%%%%%%%%%%%%%%%%%%%%%%%%%%%%%%%%%%%%%%
%%%%%%%%%%%%%%%%%%%%%%%%%%%%%%%%%%%%%%%%%%%%%%%%%%%%%%%%%%%%%%%%%%%%%

Due to the cyclic nature of the construction, the subtree $\rsymb$, $s\ge
{2^{k-1}-2}$ is, in general, identical to all subtrees
$\rsymb[s+(2^k-1)\ell']$, $\ell'
\ge 0$, up to appropriate scaling by $q^{(2^k-1)\ell'}$. In the example of
Figure~\ref{fig:code_1_8}, the tree $\rsymb[9]$ is identical to the tree
$\rsymb[2]$, indicated in the figure as $\rsymb[2+7\ell']$. An additional
source of self-similarity is provided by the trees $\Ltreek$ and $\Ltreem$;
in Figure~\ref{fig:code_1_8}, the sub-tree labeled $q^{10}\Ltreem[3]$ is
identical to that labeled $q^{9}\Ltreem[3]$, etc. Overall, although the width
of the tree is unbounded (driven by the $\ell$ copies of $\macrosymbol$ in
each layer of Case 2), the total number of distinct sub-trees in $\Cminus$ is
finite.

The following theorem enumerates the code lengths assigned to
signatures by the codes $\Cminus$. It follows immediately from the
description of the codes in~(\ref{eq:case1})
and~(\ref{eq:(i)})--(\ref{eq:(v)}).

\begin{theorem}
\label{theo:s_length}
Code $\Cminus,\;k{>}1$, assigns code lengths $\Lambda_s$ or
$\Lambda_s+1$ to signatures $s$ according to the expressions for
$\Lambda_s$ and the codeword counts in Tables~\ref{tab:Case1}
and~\ref{tab:Case2}, corresponding, respectively, to the cases $0 \le s
\le 2^{k-1}-2$ (Case 1) and $s \ge 2^{k-1}-1$ (Case 2).
\end{theorem}

We now present some auxiliary results that will be useful in proving the
optimality of the codes $\Cminus$. We rely on the following relations, which
are readily derived from the definitions of the respective trees, under the
assumption $q=2^{-k}\,$:
\begin{equation}\label{eq:sweights}
w(\Ctree_k) = 2w(\Ctree_{k-1}) = w(\Ltreek) = 2w(\Ltreem) = q^{-1}\,.
\end{equation}

The next lemma bounds the weight of the symbol $\rsymb$
in~(\ref{eq:case1}) and~(\ref{eq:(i)})--(\ref{eq:(v)}).

\begin{lemma}\label{lem:weight(R_s)}
When $s\le 2^{k-1}-2$ (Case 1), we have $0 \le w(\rsymb) \le
\frac{7}{9}$. When $s >2^{k-1}-2$ (Case 2), we have $\half \le
w(\rsymb) \le 1$.
\end{lemma}

%%%%%%%%%%%%%%%%%%%%%%%%%%%%%%%%%%%%%%%%%%%%%%%%%%%%%%%%%%%%%%%%%%%%%%%%%%%%%
% Following two tables should go together!
% Placement is tricky
%%%%%%%%%%%%%%%%%%%%%%%%%%%%%%%%%%%%%%%%%%%%%%%%%%%%%%%%%%%%%%%%%%%%%%%%%%%%%
\newlength{\tmpfloatsep}
\newlength{\tmptextfloatsep}
\setlength{\tmpfloatsep}{\floatsep}
\setlength{\tmptextfloatsep}{\textfloatsep}
\setlength{\floatsep}{0.5\floatsep}
\setlength{\textfloatsep}{0.5\textfloatsep}
\begin{table}[t]
\caption{\label{tab:Case1}Code lengths and codeword counts for codes $\Cminus$ on signatures $s$,
$0 \le s\le 2^{k-1}-2$.}
\begin{center}
\renewcommand{\arraystretch}{1.2}
\begin{tabular}{|p{9.5em}|p{11em}|p{5.5em}|}
\hline
\multicolumn{3}{|l|}{$
\begin{array}{ll}
\text{\bf Case 1:} &  0 \le s \le 2^{k-1}-2,\;\;\;s=2^i+j-1,\;\;\;\;0\le i \le k{-}2    \\[-0.75ex]
               & \Lambda_s=(s+2)(i+1) - 2^{i+1}
 \end{array}
 $ %
 }  \\
\hline
 & \multicolumn{2}{c|}{ $\quad\quad$ {\bf Number of codewords (signatures)}}\\
 \cline{2-3}
{\bf Range of $j$ } &\multicolumn{1}{c|}{ length $\Lambda_s$ }& \multicolumn{1}{c|}{ length $\Lambda_s{+}1$}\\
\hline
$0\le j \le 2^i-1$& $(2^i-j-1)$& $2j+1$ \\
\hline
\end{tabular}
\end{center}
\end{table}
\begin{table}[t]
\caption{\label{tab:Case2}Code lengths and codeword counts for codes $\Cminus$ on signatures  $s
\ge 2^{k-1}-1$.}
\begin{center}
\renewcommand{\arraystretch}{1.2}
\begin{tabular}{|p{9.5em}|p{11em}|p{5.5em}|}
\hline
\multicolumn{3}{|l|}{$
\begin{array}{ll}
\text{\bf Case 2:} &  s \ge 2^{k-1}{-}1,\;\;s = 2^{k-1}{-}1{+}(2^k{-}1)\ell{+}j,
\;\;\ell\ge 0 \\[-0.75ex]
               & \Lambda_s = (s+2)k -2^k
 \end{array}
 $ %
 }  \\
\hline
 & \multicolumn{2}{c|}{ $\quad\quad$ {\bf Number of codewords (signatures)}}\\
 \cline{2-3}
{\bf Range of $j$ } &\multicolumn{1}{c|}{ length $\Lambda_s$ }& \multicolumn{1}{c|}{ length $\Lambda_s{+}1$}\\
\hline
$0\le j\le 2^{k-1}{-}3$ &  $(2^k{-}1)\ell + (2^{k-1}{-}j{-}1)$ & $ 2^j{+}1 $\\
 $j = 2^{k-1}{-}2$ &   $ (2^k{-}1)\ell$ & $ 2^k{-}2$ \\
 $2^{k-1}{-}1\le j\le 2^k{-}4$ & $ (2^k{-}1)\ell + 3{\cdot}2^{k-1}
{-}2{-}j$ & $ 2j{+}2{-}2^k$\\
$j = 2^k{-}3$ & $(2^k{-}1)\ell + 2^{k-1}{+}1$ & $ 2^k{-}4$ \\

$ j = 2^k{-}2$ & $ (2^k{-}1)\ell{+}2^{k-1}{-}1 $ & $ 2^k{-}1$ \\
\hline
\end{tabular}
\end{center}
\end{table}
\setlength{\floatsep}{\tmpfloatsep}
\setlength{\textfloatsep}{\tmptextfloatsep}
%%%%%%%%%%%%%%%%%%%%%%%%%%%%%%%%%%%%%%%%%%%%%%%%%%%%%%%%%%%%%%%%%%%%
%%%%%%%%%%%%%%%%%%%%%%%%%%%%%%%%%%%%%%%%%%%%%%%%%%%%%%%%%%%%%%%%%%%%

\begin{IEEEproof}
For $s\le 2^{k-1}-2$, we have
\begin{align}
w(\rsymb)  &=
\sum_{s'=s+1}^{\infty}(s'+1)q^{-s}
w(s')\nonumber\\
& =
\sum_{r=0}^{\infty}
(s+r+2)q^{r+1}
=
\frac{(s+1)(1-q)+1}{(1-q)^2}\,q\,.\label{eq:wRs1}
\end{align}
The right-hand side of~(\ref{eq:wRs1}) increases with $s$. Setting
$s=2^{k-1}-2 =
\frac{1}{2q}-2$,
we obtain $\weight{\rsymb} =
\half\left(1+\frac{q(1+q)}{(1-q)^2}\right)$,
which satisfies the claimed upper bound for $q \le
\frac{1}{4}$.
When $s\ge 2^{k-1}-1$, $\rsymb$ contains all the signatures $s' > s$
(with their weights scaled by $q^{-s}$) that are not contained in the
components $q\Ltreek$ of the groups $\macrosymbol$, or in a possible
sibling $q\,\Ctree_{k-1}$ or $q\Ltreem$ of $\rsymb$. Write  $s$ as
in~(\ref{eq:s}). The scaled total weight of signatures $s'>s$ is
\begin{align*}%label{eq:totw}
W_{s} &=
q^{-s}
\sum_{r=0}^{\infty}
(s+2+r)q^{s+1+r}=\frac{(s+2)q}{1-q}+\frac{q^2}{(1-q)^2}
\nonumber\\
&=
\frac{2q(1+j)+1}{2(1-q)}
+
\frac{q^2}{(1-q)^2}+
\ell \,,
\end{align*}
where the last equality follows by applying~(\ref{eq:s}) and
substituting $q^{-1}$ for $2^k$. Let $W'_s$ denote the part of $W_s$
that is contained in the symbols $q\Ltreek$, $q\,\Ctree_{k-1}$, or
$q\Ltreem$ mentioned above. Observing the layer structures
in~(\ref{eq:(i)})--(\ref{eq:(v)}), and applying~(\ref{eq:sweights}),
we obtain $W'_s =\ell +\delta$, where:
\begin{equation}\label{eq:delta}
\delta =
\left\{ \,
\begin{array}{ll}
0, & 0 \le j \le 2^{k-1}-3,\\
\half,& 2^{k-1}-2 \le j \le 2^k-3,\\
1,& j = 2^k-2\,.
\end{array}
\right.
\end{equation}
The claim of the lemma for $s > 2^{k-1}-2$ follows by writing
$\weight{\rsymb}=W_s - W'_s$, observing that $\weight{\rsymb}$
increases monotonically with $j$, and bounding $\weight{\rsymb}$, as an
elementary function of $q$ , in the interval $0 < q\le\frac{1}{4}$ for
each of the cases in~(\ref{eq:delta}). Notice that due to the mentioned
monotonicity, $\weight{\rsymb}$ is evaluated only at the ends of the
ranges of $j$ in~(\ref{eq:delta}), and we substitute $q^{-1}$ for
$2^k$.
\end{IEEEproof}
The following is an immediate consequence of
Lemma~\ref{lem:weight(R_s)}.
\begin{corollary}\label{cor:sorted}
Let $\rsymb'$ denote the virtual symbol containing $\rsymb$ in each
layer $\layer$ listed in~(\ref{eq:case1})
and~(\ref{eq:(i)})--(\ref{eq:(v)}).  Then, after scaling by $q^{-s}$,
all the symbols to the left of $\rsymb'$ in $\layer$ are of weight
$1$,
all the symbols to its right are of weight $2$, and we have $1 \le
\weight{\rsymb'} \le 2$.
\end{corollary}
\begin{IEEEproof}
The claims on the symbols to the left and to the right of $\rsymb'$
follow from~(\ref{eq:sweights}) and the definition of the notation
$\macrosymbol$ in~(\ref{eq:macro}). As for $\rsymb'$, we have
$\weight{\rsymb'} = 1+\weight{\rsymb}$, and the claim of the corollary
follows by applying Lemma~\ref{lem:weight(R_s)}.
\end{IEEEproof}

\begin{theorem}\label{theo:q<1/2}
The prefix code $\Cminus$ is optimal for $\tdgd(q)$ with $q=\qparI$,
$k>1$.
\end{theorem}
\begin{IEEEproof}
As before, we rely on the method from~\cite{gv75}. The reduced sources are
defined by $\mathcal{S}_s = \mathcal{H}_s \cup \mathcal{F}_s $, where
$\mathcal{H}_s$ denotes, as before, the multiset of signatures strictly
smaller than $s$, and the multiset $\mathcal{F}_s$ is essentially identical
to the layer $\layer$  defined in~(\ref{eq:case1})
and~(\ref{eq:(i)})--(\ref{eq:(v)}). The steps taking a reduced source to one
of lower order follow the layer ``unfolding'' steps listed in the description
of the codes $\Cminus$ (see the discussion following~(\ref{eq:case1})
and~(\ref{eq:(i)})--(\ref{eq:(v)}), and Appendix~\ref{app:layers}), in
reverse order (bottom-up). It remains to show that these steps correspond to
a valid sequence of mergers in the Huffman procedure. Consider a layer
$\layer$, and let $\symb_1,\symb_2,\ldots,\symb_N$ denote its symbols, listed
from left to right, as shown in~(\ref{eq:case1})
and~(\ref{eq:(i)})--(\ref{eq:(v)}).
 It is readily verified that $N=2^i$ for a layer~(\ref{eq:case1}),
with $i$ as defined in Case 1, and that $N$ is divisible by $2^{k-1}$ in
layers of type (i)--(ii), and by $2^k$ in layers of type (iii)--(v). %
By Corollary~\ref{cor:sorted}, the $\symb_j$ are ordered by increasing weight
order, and, since $q<1/2$, the weight of any $\symb_j$ is smaller than any
weight in $\mathcal{H}_s$. Thus, the Huffman procedure on $\mathcal{S}_s$
starts by pairing symbols in $\layer$. Now, it also follows from
Corollary~\ref{cor:sorted} that the merger of any two of the $\symb_j$
results in a combined weight that is at least as large as any weight in the
layer. Thus, merging $\symb_{2j-1}$ with $\symb_{2j}$, $1\le j\le N/2$, is a
valid sequence of steps in the Huffman procedure on $\layer$. Moreover, since
there is at most one symbol of weight different from $1$ or $2$ (after
scaling), and strictly between them, the resulting sequence of merged weights
includes weights $2$, $\omega$, and $4$, with $2 \le \omega \le 4$, with at
most one symbol of weight $\omega$. We iterate the argument until the
signatures $s{-}1$ get incorporated, and $\layer[s-1]$ gets formed (see
Appendix~\ref{app:layers}), reaching, thus, the reduced source
$\mathcal{S}_{s-1}$.
Proceeding recursively, we reach the reduced source $\mathcal{S}_0$, which coincides with the layer $\layer[0]$. As described in~(\ref{eq:case1}) for $s=0$, this layer consists of one virtual symbol formed by $\mathcal{R}_0$ and the symbol $0$ joined under the root of the tree $\Cminus$ (thus, the Huffman procedure on $\mathcal{S}_0$ is trivial in this case).
\end{IEEEproof}

%%%%%%%%%%%%%%%%%%%%%%%%%%%%%%%%%%%%%%%%%%%%%%%%%%%%%%%%%%%%%%%%%%%%%%
%%%%%%%%%%%%%%%%%%%%%%%%%%%%%%%%%%%%%%%%%%%%%%%%%%%%%%%%%%%%%%%%%%%%%%

\subsection{A limit code}
The sequence of optimal codes $\Cminus$  stabilizes in the limit of
$k\to\infty$ ($q\to 0$), as stated in the following corollary.
\begin{figure}
\centering{\includegraphics[width=2.5in]{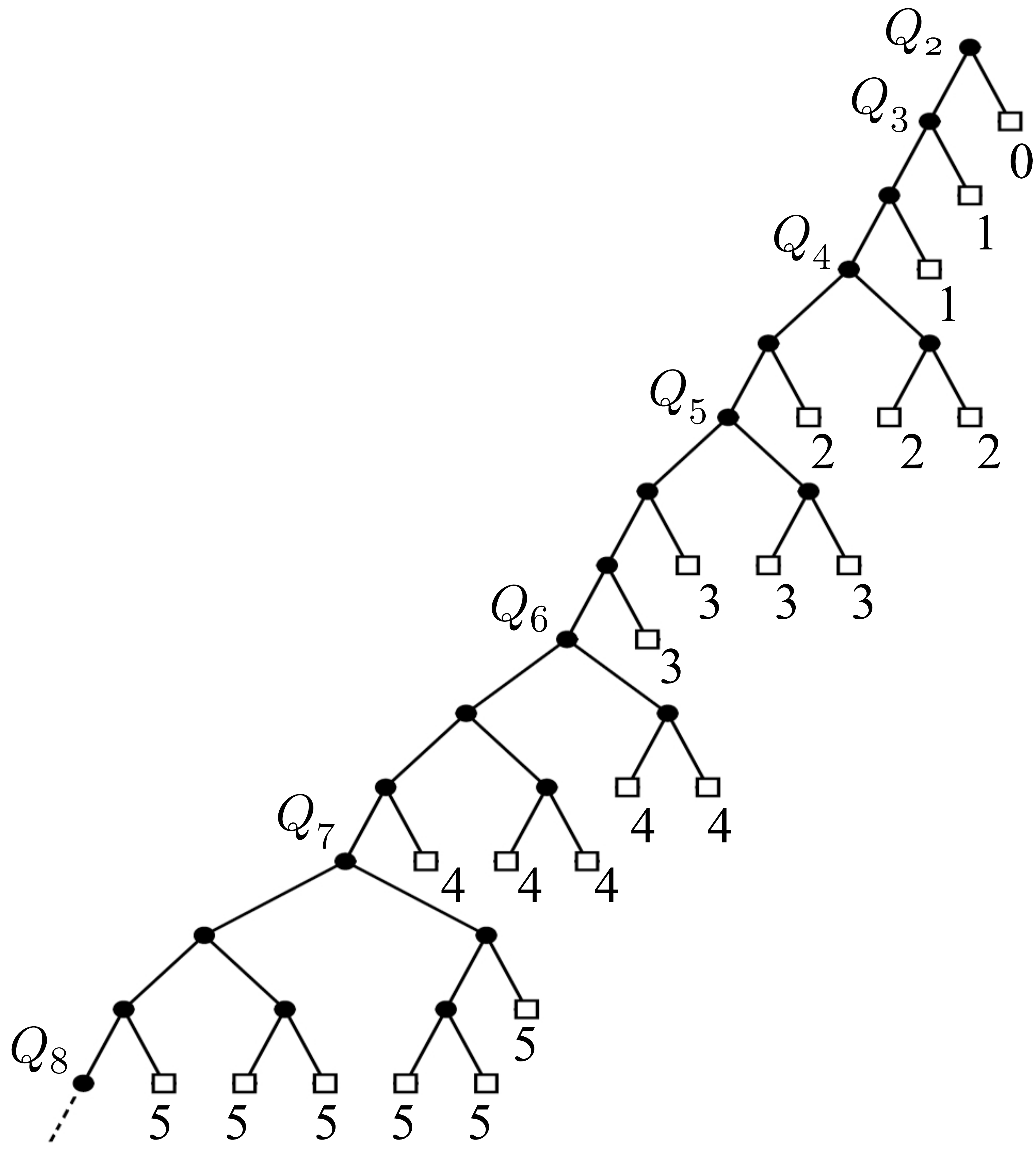}}
{\caption{\label{fig:limitnormalcode} Top of the limit tree
$\limcode$.} }
\end{figure}
\begin{corollary}\label{theo:limit}
When $k{\to}\infty$, the sequence of optimal trees $\Cminus$
converges to a limit tree $\limcode$ that can be constructed as
follows: start with $Q_n$ for $n{=}2$, recursively replace the
leftmost leaf of the deepest level of the current tree by $Q_{n+1}$, and increase~$n$.
\end{corollary}
\begin{IEEEproof} The corollary is proved by observing that the part of
the tree corresponding to $0 \le s \le 2^{k-1}$ in
Theorem~\ref{theo:q<1/2} remains invariant for all $k'\ge k$. This
corresponds to the layers $\layer$ of Case 1.
\end{IEEEproof}

The limiting property of $\limcode$ in connection with the $\tdgd$ is
mentioned also in~\cite[Ch.\ 5]{MBaer2003}.
Figure~\ref{fig:limitnormalcode} shows the first fourteen levels of
$\limcode$. Notice that the first eleven levels coincide with those of
$\Cminus[3]$ in Figure~\ref{fig:code_1_8}, up to reordering of nodes at
each level. Explicit encoding with $\limcode$ can be done as follows.
Given a pair $(i,j)$, with signature $s=i{+}j$, we write $s = 2^t-1+r$,
with $0
\le r \le 2^t-1$ and $t\ge 0$. We encode $(i,j)$ with a binary codeword
$xy$, where $x=1^{(t-1)(s+1)+2r+1}$ identifies the path to the root of
the quasi-uniform tree that contains all the leaves of signature $s$,
and $y=Q_{s+2}(i+1)$.  The resulting code length distribution for
signature $s$ is: $2^t-1-r$ signatures encoded with length
$(t-1)(s+2)+2r+2$, $2r+1$ signatures encoded with length
$(t-1)(s+2)+2r+3$.

The following corollary shows the average code length attained by
$\limcode$ on an arbitrary TDGD.
\begin{corollary}
The average code length  of the limit code $\limcode$ under
$\tdgd(q)$ is given by
\[
\limlen= 1+\frac{1}{1-q}\sum_{t\geq 0}q^{2^t}(2^t(1-q)+2)\,.
\]
\end{corollary}

\begin{IEEEproof}
\nopagebreak
For $s\ge 0$, let $r$ and $t$, $t\ge 0$,  $0\le r
\le 2^t-1$, be the (uniquely determined) integers such that  $s = 2^t-1+r$.
By Corollary~\ref{theo:limit} and the ensuing discussion,
we can write
\begin{equation}
\ncostt_q(\limcode)
=(1-q)^2\sum_{t\geq 0}
\sum_{s=2^t-1}^{2^{t+1}-2}
 q^s \DDD(t,s) \label{eq:limlength_line1}\,,
\end{equation}
where
\[
\DDD(t,s) =
\left(\,\rule{0pt}{0.9em}(t-1)(s+2)+2r+2\,\right)(s+1)+2r+1\,.
\]
Substituting $r=s-2^t+1$ and carrying out the inner summation
in~(\ref{eq:limlength_line1}), we obtain
\begin{align}
\ncostt_q(\limcode) =& (1-q)^2\sum_{t\geq 0} \left ( q^{2^{t+1}-1}A(t) +
q^{2^{t}-1}B(t)
\right)\,,\label{eq:limlength}
\end{align}
for some functions $A(t)$ and $B(t)$. It can be verified by symbolic
manipulation that
\[
B(0)=\frac{1-q^2+2q}{(1-q)^3}
\]
and
\[
 A(t-1)+B(t)=q\frac{2^t-2^tq+2}{(1-q)^3}\,.
\]
 Substituting in~(\ref{eq:limlength}), after rearranging terms, we obtain
\begin{align*}\squeeze
\limlen&=(1{-}q)^2\Biggl( B(0){+}\sum_{t\ge 1} q^{2^t-1}\Big(A(t{-}1){+}
B(t)\Big)\Biggr)\\
&= (1{-}q)^2\Biggl(\frac{1{-}q^2{+}2q}{(1-q)^3}
+\sum_{t\geq 1}q^{2^t}\frac{2^t{-}2^tq{+}2}{(1-q)^3}\Biggr)\\
&=1+\frac{1}{1-q}\sum_{t\geq
0}q^{2^t}(2^t(1-q)+2)\,.
\end{align*}
\end{IEEEproof}

\section{Practical considerations and redundancy}
\label{sec:plots}

\begin{figure*}
\begin{center}
\includegraphics[width=0.98\linewidth]{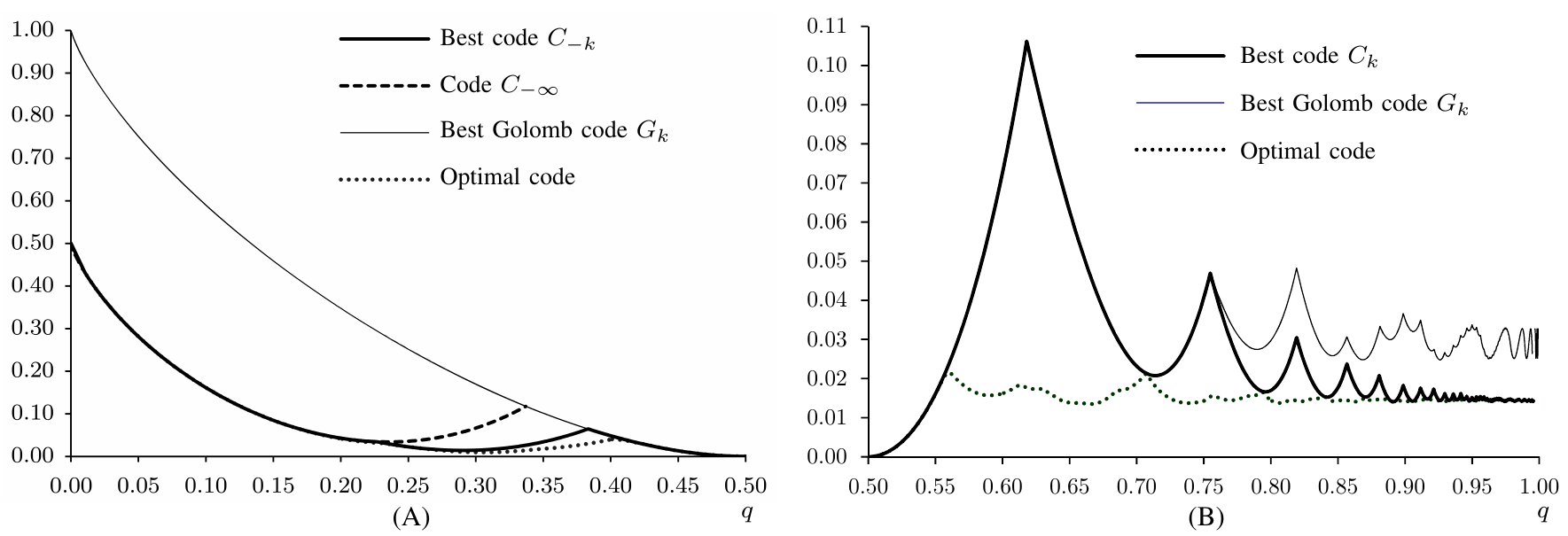}
\end{center}
\caption{\label{fig:plots}Redundancy (in bits/integer symbol) for the optimal
prefix
  code (estimated numerically), the best Golomb code,
  the limit code $\limcode$, and the best code $\Cminus$ or $C_k$ for
each value of $q$,
  (A) $0<q<\frac{1}{2}$, (B) $\frac{1}{2}\leq q<1$.
  The limit code $\limcode$ is plotted up to $q=0.33715\ldots$,
  where its curve intersects that of $C_1$ (or, equivalently, $C_{-1}$).}
\end{figure*}

In a practical situation, one could use the codes $\Cplus$ for  $q\ge
\half$, and the codes $\Cminus$ for $q<\half$. However, a lower
complexity alternative, which incurs a modest code length penalty (as
shown in Figure~\ref{fig:plots}), is to  use  $\limcode$ in lieu of the
codes $\Cminus$, up to the value of $q$ where switching to $\Cplus[1]$
gives better average code length. The crossover point is at $q \approx
0.33715$.

Encoding a symbol pair $(x,y)$ with a code $\Cplus$ is of about the
same complexity as two encodings of individual symbols with a Golomb
code of order $k$. As described in Theorem~\ref{theo:opt-code}, the
encoding with $\Cplus$ entails unary encodings of $\lfloor x /
k\rfloor$ and $\lfloor y /k\rfloor$, which would also be needed with
the Golomb code. Given the profile of the top code $T_k=\Tskck$,
determined in Theorem~\ref{theo:top}, encoding with $T_k$ requires
comparing the index of the pair $(x \bmod k, y \bmod k)$ with at most
two fixed thresholds, to determine the corresponding code length (which
can assume up to three consecutive integer values). The codeword is
then computed directly from the index. Each encoding with the Golomb
code, on the other hand, requires one comparison with a fixed threshold
to determine the code length of each $Q_k$ component, or a total of two
for the pair $(x,y)$.

As in the one-dimensional case (see, e.g.,~\cite{msw00},
\cite{SW97}), when encoding a sequence $x_1,x_2,\ldots,x_{2t},\ldots$,
the best code for the next pair $(x_{2t-1},x_{2t})$ can be determined
adaptively, driven by the sufficient statistic $S_t =
t^{-1}\sum_{j=1}^{2t-2} x_j$. The crossover points for the estimates of
the code parameter $k$ can be precomputed and stored in terms of the
statistic $S_t$. The one-dimensional code has a slight advantage in the
adaptation, in that it can adapt its statistic with every symbol,
whereas the two-dimensional code can only do it every two symbols.
Depending on the application, this advantage is likely to be superseded
by the redundancy advantage of the two-dimensional code. Also as in the
one-dimensional case, there are certain complexity advantages, in both
encoding and adaptation when using the subset of parameters of the form
$k=2^r$. In this case, an adaptation strategy that estimates the best
parameter $r$ directly from the statistic $S_t$, without the need to
compare it with precomputed crossover points, can be derived for the
codes $\Cplus$, as was done in~\cite{msw00} and~\cite{SW97} for
two-sided geometric distributions. We omit the details, since both the
technique and the resulting parameter estimation method are similar to
those in the references.

\begin{figure*}
\begin{center}
\includegraphics[width=0.98\linewidth]{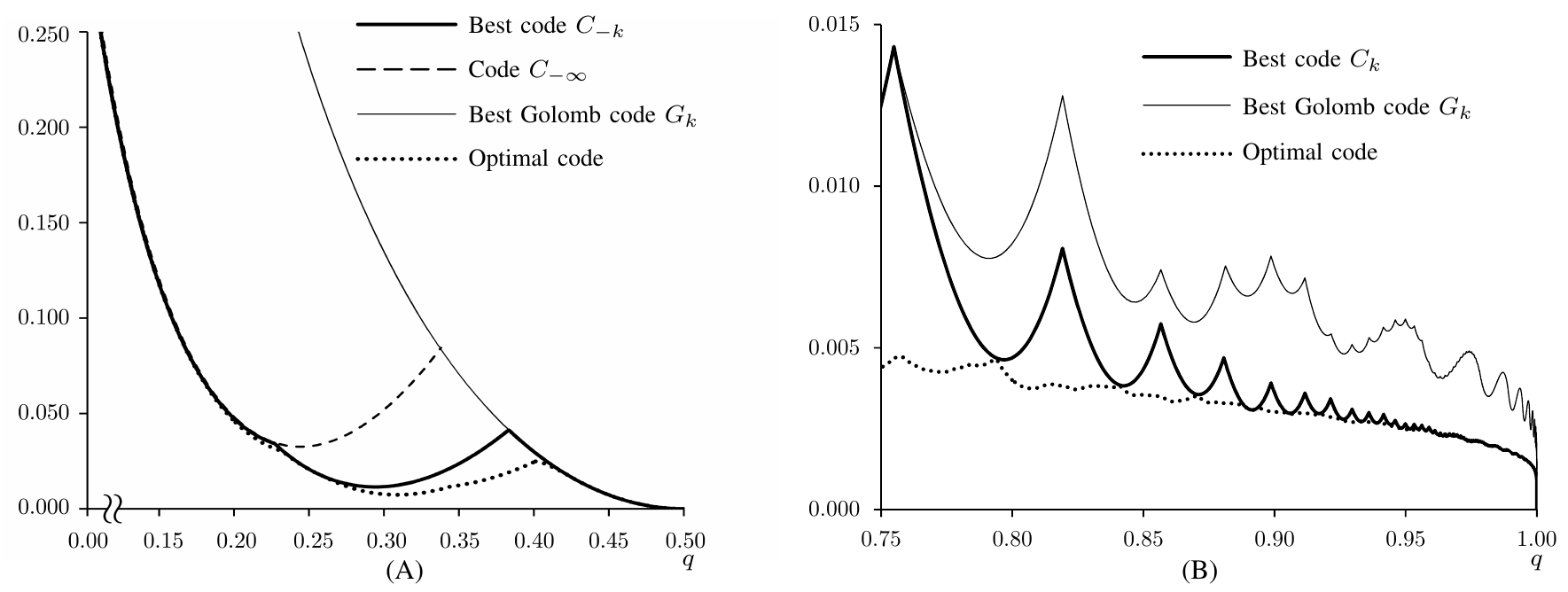}
\end{center}
\caption{\label{fig:plots-rel}Relative redundancy (redundancy normalized by the per-symbol entropy)
for the codes of Figure~\ref{fig:plots}. The interval $0.5 \le q < 0.75$
is omitted from (B), as the best codes $C_k$ and $G_k$ coincide over that interval. }
\end{figure*}

Figure~\ref{fig:plots} presents plots of redundancy for various code
families as a function of $q$, measured in bits per integer symbol
relative to the entropy of the geometric distribution (recall that the
latter is given by $H(q) =
\frac{h(q)}{1-q}$, where $h(q)$ is the binary entropy function~\cite{gv75}). Plots
are shown for the optimal prefix code for each value of $q$ (estimated
numerically over a dense grid of values of $q$, and in sufficient
precision to make the estimation error smaller than the plot
resolution), the best Golomb code, the best code $\Cminus$ or $\Cplus$
for each $q$, and the limit code $\limcode$. Here, ``the best Golomb
code'' means the code $G_k$ that minimizes (over $k$) the code length
for the given value of $q$; similar minimizations are used for the best
codes $\Cminus$ and $\Cplus$ for each $q$. In the figure, we can
observe the advantage in redundancy for the codes $\Cminus$ (or
$\limtree$) and $\Cplus$ over Golomb codes, except in the region where
the best codes of both types are equivalent (i.e., the optimality
regions of $\Cplus[1]$ and $\Cplus[2]$). The redundancy advantage is
near $2:1$ (as expected) at the limit of $q\to 0$ and it peaks near
$q=0.28$ (at more than $13.6:1$). A redundancy advantage close to $2:1$
is observed also as $q\to 1$. The advantage of $\Cplus$ over
symbol-by-symbol Golomb codes is consistent with
Corollary~\ref{cor:golomb-not-optimal}, and, in fact, the plot in
Figure~\ref{fig:plots} can be regarded as ``visual evidence'' for the
corollary. Figure~\ref{fig:plots-rel} plots the corresponding curves
for the \emph{relative redundancy}, i.e., the redundancy normalized by
the per-symbol entropy $H(q)$ for each plotted value of $q$. We observe
that although the relative redundancy for all the codes considered
converges to zero, as expected, when $q\to 1$ (since $H(q)\to\infty$),
the decay is very slow for most of the interval, and the curves fall to
zero ``suddenly'', with infinite slope, near $q=1$. This is due to the
slow rate of growth of $H(q)$, which behaves asymptotically as
$-\log(1-q)$ near the limit point.

It is apparent from Figure~\ref{fig:plots} that as the redundancy of
the codes $C_k$ peaks in the transitions between one ``best'' value of
$k$ and the next, the estimated redundancy of the optimal codes remains
rather flat. This poses the question, which also remains open, of
whether other sequences of codes with simple descriptions and
encoding/decoding procedures could be found, that would more closely
track the redundancy curve of the optimal codes.

\begin{figure}
\begin{center}
\includegraphics[width=0.90\linewidth]{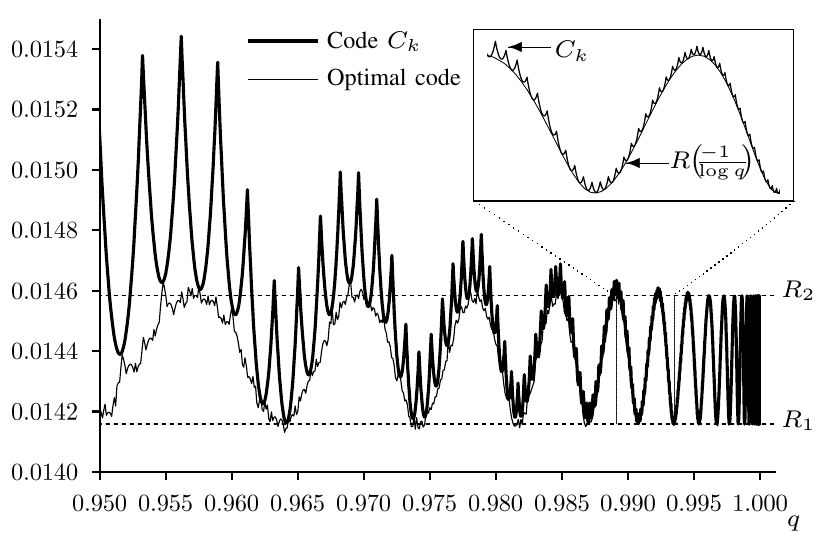}
\end{center}
\caption{\label{fig:oscil}Redundancy as $q{\to}1\;(k{\to}\infty)$.
Dashed lines show
 the asymptotic limits $R_1$ and $R_2$. The inset closes up further on
a narrow segment, showing the redundancy of the codes $\Cplus$ vs.\
 the asymptotic estimate~(\ref{eq:R}).}
\end{figure}

The asymptotic behavior of the redundancy of $\Cplus$ in the regime
$q\to 1$, shown in more detail in Figure~\ref{fig:oscil}, is
oscillatory, as is also the case for Golomb codes~\cite{gv75}. The
limiting behavior of the redundancy can be characterized precisely, as
we show next.

\begin{corollary}\label{cor:oscil} Let $\lambdak = 2^M/k^2$, where $M$
is as defined in Theorem~\ref{theo:top}. As $k\to\infty$, the
redundancy of the code $\Cplus$ at $q=\qparII$ is
\begin{align}
R(k) = & \half\left(1+\log \lambdak\right){+}
{2}^{1-2\sqrt {{\lambdak}-\half}}
 \left( 1{+}\frac{2}{\log e}\sqrt {{\lambdak}-\half} \right)\nonumber\\
& - \log(e\log e)+ o(1)\,.\label{eq:R}
\end{align}
\end{corollary}

\textbf{Remark.} We have $\frac{3}{4} \lessapprox \lambdak
\lessapprox \frac{3}{2} $, where $\lessapprox$ denotes inequality up
to asymptotically negligible terms. For large $k$, as $k$ increases,
$\lambdak$ sweeps its range decreasing from $\frac{3}{2}$ to
$\frac{3}{4}$, at which point $\Mk$ increases by one, and $\lambdak$
resets to $\frac{3}{2}$, starting a new cycle.
\begin{IEEEproof}[Proof of Corollary~\ref{cor:oscil}] We derive,
  from~(\ref{eq:avglenCkqk}), an asymptotic expression for the code
  length $\ncost{\Cplus}$. To estimate the parameter $j$
  in~(\ref{eq:avglenCkqk}), we need to solve the quadratic equation
  $\fxdelta(x)=0$, with $\fxdelta(x)$ as defined in
  Theorem~\ref{theo:top}. Writing $2^M = \lambdak k^2$, it is readily
  verified that the largest solution to the equation is $\xi
  =\left(2\sqrt{\lambdak-\half}-1\right)\,k +O(1) \defined \alpha\,k +
  O(1)$. Thus, $j = \alpha\,k+O(1)$, and $q^j=2^{-\alpha}+O(k^{-1})$.
  Writing also $q=\qparII = 1-\frac{\ln 2}{k} + O(k^{-2})$, and noting
  that $\fxdelta(j)= O(k)$, we obtain, from~(\ref{eq:avglenCkqk}),
\[
\ncost{\Cplus} = M + 1 + 2^{1-\alpha}\big(\,1 + (1+\alpha)\ln 2\big) +
o(1)\,.
\]
As for the entropy, we have
\begin{align*}
H(q) &= \frac{-q\log q}{1-q} - \log(1-q)
= \log(e\log e) + \log k + o(1)\\
 &=  \log(e\log e) + \half\left( M -
\log\lambdak\right)+o(1)\,.
\end{align*}
The claimed result~(\ref{eq:R}) follows by substituting the asymptotic
expressions for $\ncost{\Cplus}$ and $H(q)$ in the formula for  the
redundancy per symbol, namely, $R(k) = \half\ncost{\Cplus} - H(q)$.
\end{IEEEproof}

The limits of oscillation of the function $R_k$ can be obtained by
numerical computation, yielding  $R_1 \defined \liminf_{k\to\infty}
R(k) = 0.014159{\ldots}\,$ and $R_2 \defined \limsup_{k\to\infty} R(k)
= 0.014583{\ldots}\,$. These limits are shown in
Figure~\ref{fig:oscil}. The corresponding limits for the redundancy of
the Golomb codes are, respectively, $R_1'=0.025101{\ldots}\,$ and
$R_2'= 0.032734{\ldots}\,$~\cite{gv75}.

Corollary~\ref{cor:oscil} applies to the discrete sequence of
redundancy values at the points $q=\qparII$. It is not difficult to
prove that the same behavior, and in particular the limits $R_1$ and
$R_2$,  apply also to the continuous redundancy curve obtained when
using the best code $\Cplus$ at each arbitrary value of $q$. This
follows from the readily verifiable fact that as $q$ varies in the
interval $2^{-1/k} \le q \le 2^{-1/(k+1)}$, the maximal variation in
both the code length under $\Cplus$ and the distribution entropy is
bounded by $O(k^{-1})$. Figure~\ref{fig:oscil} suggests that the same
oscillatory behavior might apply also to the redundancy curve of the
optimal prefix code for each value of $q$. It follows from the
foregoing discussion that this is true for the limit superior $R_2$.
The question remains open, however, for the limit inferior $R_1$, which
is an upper bound for the limit inferior
of the optimal redundancy.

\appendices

\section{Proofs for Subsection~\ref{sec:fringe2}}
\label{app:fringe}

We recall that we consider a $4$-uniform probability distribution
$\bldp=(p_1,p_2,\ldots,p_N)$, where probabilities are listed in
non-increasing order, and an optimal tree $T$ for $\bldp$, with $\fT
\le 2$. We define $m=\lceil\log N\rceil$, and we denote by $n_\ell$
the number of leaves at depth $\ell$ in $T$.

\begin{IEEEproof}[Proof of Lemma~\ref{lem:claimA}]
Say $T$ has $t>0$ leaves at depths $\ell<m{-}2$. Then, $T$ has no
leaves at depths $\ell'\ge m$, and it can have a total of at most
$2^{m-1}-3t$ leaves altogether. But $N>2^{m-1}$, a contradiction. Say
now that $T$ has nodes at depth $m{+}2$. Then all of its leaves must be
at depths $\ell'\ge m$, and some must be at depths strictly greater
than $m$. Thus, $T$, being full, must have more than $2^m\geq N$
leaves, again a contradiction. The second claim of the lemma is a
straightforward consequence of $\fringeT\leq 2$.
\end{IEEEproof}

\begin{IEEEproof}[Proof of Lemma~\ref{prop:profiles1}]
Let $\nT=(n_{M-1},n_{M},n_{M+1})$ be the compact profile of a tree $T$
with $N$ leaves and $\fT\le 2$.  Clearly, $n_{M+1}$ must be even, and
we write $n_{M+1} = 2c$ for some nonnegative integer $c$. The
components of $\nT$ must satisfy
\begin{equation}\label{eq:sumn}
n_{M-1} + n_{M} + 2c = N\,.
\end{equation}
By Kraft's equality, which must hold for the full tree $T$, we have
\begin{equation}\label{eq:kraft2}
4 n_{M-1} + 2 n_{M} + 2c = 2^{M+1} \,,
\end{equation}
which holds also in the case $c=0$. From~(\ref{eq:sumn})
and~(\ref{eq:kraft2}), we obtain
\begin{equation}\label{eq:nm1B}\squeeze
n_{M-1} = 2^M - N + c\,.
\end{equation}
Now, from~(\ref{eq:nm1B}) and~(\ref{eq:sumn}), we obtain
\begin{equation}\label{eq:nm}\squeeze
n_M =  2N - 2^M - 3c\,.
\end{equation}
Equations~(\ref{eq:nm1B}) and~(\ref{eq:nm}), together with the
definition of $c\,$ yield the profile~(\ref{eq:cprofile}). The valid
range of variation of $c$ is determined by the non-negativity
constraints on the entries of the profile. When $M=m-1$ ($\short=1$),
the lower limit $\ccmin=N-2^{m-1}$ is determined by the nonnegativity
of $n_{M-1}$. Since $2^M \ge N$ when $M=m$, the lower limit is the
trivial $\ccmin[0] = 0$ in this case. In both cases, the upper limit
$\ccmax =
\lceil\frac{2N-2^M}{3}\rceil$ is determined by the nonnegativity of
$n_M$.
\end{IEEEproof}

\begin{IEEEproof}[Proof of Lemma~\ref{lem:cascade}]
For a given value of $\short\in\cerouno$, assume $c$ and $c'$ are
indices such that $\ccmin < c' \le c \le \ccmax$, and let
$\sseq_\short$ be the segment of $\sseq$ corresponding to $\short$.
By~(\ref{eq:Dc}) and the monotonicity of the weights, we have
\begin{align*}
\DD[c'] & = p_{N-2c'+1} + p_{N-2c'+2} - p _{2^M-N+c'} \\
&\le
p_{N-2c+1} + p_{N-2c+2} - p_{2^M-N+c} = \DD\,.
\end{align*}
Thus, if $\DD<0$ then $\DD[c']<0$, and if $\DD=0$ then $\DD[c']\le 0$.
It follows that $\sseq_\short$ is non-decreasing. It remains to prove
that $-\sg(\DDsc[1,\,{\ccmin[1]+1}]) \le \sg(\DDsc[0,1])$. Assume that
$\DDsc[0,1]\le 0$. Then, we have
\begin{eqnarray*}
\DDsc[1,\,{\ccmin[1]+1}] &=& p_{2^m-N-1} + p_{2^m-N} - p_1
 \ge 2p_{2^m-N+1} - p_1\nonumber\\
  &\ge& 2\left(p_{N-1}+p_N\right)-p_1  \ge 4p_N - p_1 \ge 0\,,
\end{eqnarray*}
where the equality follows from~(\ref{eq:Dc}) and the definition of
$\ccmin[1]$, the first and third inequalities from the monotonicity of
$\bldp$, the second inequality from our assumption on $\DDsc[0,1]$, and
the last inequality from the 4-uniformity of $\bldp$. Hence, we must
have $\DDsc[1,{\ccmin[1]+1}]\ge 0$. Similarly, if $\DDsc[0,1]<0$, then
we must have $\DDsc[1,{\ccmin[1]+1}]> 0$.  Therefore,
$-\sg\left(\DDsc[1,\,{\ccmin[1]+1}]\right)
\le
\sg(\DDsc[0,1])$, as claimed.
\end{IEEEproof}

\begin{IEEEproof}[Proof of Theorem~\ref{theo:choose}]
The theorem follows directly from Lemma~\ref{lem:cascade}, observing
also that by the assumptions of the theorem, and by
Lemma~\ref{prop:profiles1}, at least one of the trees $\Tc\,$,
$(1,\ccmax[1])\preceq \pair \preceq (0,\ccmax[0])$ must be optimal for
$\bldp$.
\end{IEEEproof}

\section{Proofs for Subsection~\ref{sec:toptree}}
\label{app:proof_of_top}
We derive the proof of Theorem~\ref{theo:top} through a series of
lemmas.  We recall that we seek an optimal tree for the source $\Ak$
of~(\ref{eq:topsource}), with vector of (unnormalized) weights
\[
\bldp=(q^0,q^1,q^1,\ldots,q^j,q^j,\ldots,q^j,\ldots,q^{2k-3},q^{2k-3},q^{2k-2}),
\]
with $q=2^{-1/k}$, and where $q^j$ is repeated $j+1$ times for $0\le j
\le k{-}1$, and $2k-1-j$ times for $k \le j \le 2k{-}2$. For succinctness,
in this appendix, when we say ``optimal'' we mean ``optimal for
$\Ak$.'' Notice that, in $\bldp$, three consecutive weights are never
distinct; we refer to this fact as the ``three consecutive weights''
property. Throughout the appendix, we assume that $k>2$, as we recall
that optimal trees for $k=1,2$ are fully characterized in
Remark~\ref{rem:2afterThm2} following Theorem~\ref{theo:opt-code}.

\begin{lemma}\label{lem:cmax}
Trees $\Tc$ with $c=\ccmax$ are not optimal. Consequently, the profile
$(n_{M-1},n_M,n_{M+1})$ of an optimal tree has $n_M \ge 3$.
\end{lemma}
\begin{IEEEproof} Recalling the profile $\nT[\Tsc]$ in~(\ref{eq:cprofile}),
  with $c=\ccmax$ and $k > 2$, we have $n_M \in \{0, 1, 2\}$,
  $n_{M-1} \ge 1$ and $n_{M+1} \ge 2$. Let $q^\ell$ be the lightest
  weight on level $M-1$. By the ``three consecutive weights'' property,
  the two heaviest weights on level $M+1$ are greater than or equal to $q^{\ell+2}$.
  Recalling the expression for $\DD$ in~(\ref{eq:Dc}),
  and the interpretation that follows it, we obtain
   $\DD[\ccmax] \ge q^\ell(1-2q^2)>0$. Thus, by
   Theorem~\ref{theo:choose},
   $T_{\ccmax}$ is not optimal. An optimal tree $\Tc$ would,
   therefore, have $c<\ccmax$, and, thus, $n_M \ge 3$.
\end{IEEEproof}

The following lemma gives a first, rough approximation of the
distribution of weights by levels in an optimal tree $\Tc$, which will
allow us to identify the appropriate range (i.e.,~(\ref{eq:large})
or~(\ref{eq:small})) for the heaviest and the lightest weights on level
$M$ of the tree.

%%%%%%%%%%%%%%%%%%%%%%%%%%%%%%%%%%%%%%%%%%%%%%%%%%%%%%%%%%%%%%%%%%%%%%%%%%%%%
\begin{figure*}[t]
\begin{center}
\setlength{\unitlength}{0.085in}
\begin{picture}(53,12)(0,-4)
\put(-3.5,2){\qbezier[5](0,0)(1.25,0)(2.5,0)}
\put(-.5,2){\line(1,0){15}}
\put(21.5,2){\line(1,0){30.5}}
\put(52.5,2){\qbezier[5](0,0)(1.25,0)(2.5,0)}
\put(15.5,2){\qbezier[10](0,0)(2.5,0)(5,0)}
\put(-2,0){\makebox(0,0){$\cdots$}}
\put(3,2){\circle{1}}
\put(3,4){\makebox(0,0){$-$}}
\put(3,0){\makebox(0,0){$p_{2^M-k^2+c}$}}
\put(3,-3){\makebox(0,0){$q^{j-\eps}$}}
\put(11,2){\circle*{1}}
\put(11,4){\makebox(0,0){$-$}}
\put(3,6){\vector(1,0){8}}
\put(7,8){\makebox(0,0){decrease}}
\put(11,0){\makebox(0,0){$p_{2^M-k^2+c+1}$}}
\put(11,-3){\makebox(0,0){$q^{j}$}}
\put(18,0){\makebox(0,0){$\cdots$}}
\put(25,2){\circle*{1}}
\put(25,0){\makebox(0,0){$p_{k^2-2c-1}$}}
\put(25,-3){\makebox(0,0){$q^{2k-2-j'-\eps'}$}}
\put(29,4){\makebox(0,0){$+$}}
\put(33,2){\circle*{1}}
\put(33,0){\makebox(0,0){$p_{k^2-2c}$}}
\put(33,-3){\makebox(0,0){$q^{2k-2-j'}$}}
\put(41,2){\circle{1}}
\put(41,0){\makebox(0,0){$p_{k^2-2c+1}$}}
\put(41,-3){\makebox(0,0){$q^{2k-2-j'+\eps''}$}}
\put(49,2){\circle{1}}
\put(45,4){\makebox(0,0){$+$}}
\put(45,6){\vector(-1,0){16}}
\put(37,8){\makebox(0,0){increase}}
\put(49,0){\makebox(0,0){$p_{k^2-2c+2}$}}
\put(49,-3){\makebox(0,0){$\;\;\;\;\;\;q^{2k-2-j'+\eps''+\eps'''}$}}
\put(54,0){\makebox(0,0){$\cdots$}}
\end{picture}
\caption{\label{fig:sixweights}Weights involved in the conditions for $c=\cck$:
\ {\LARGE$\circ\,$} weights in $\DD\,$,\ {\LARGE$\bullet\,$} weights in $\DD[c+1]\,$.}
\end{center}
\end{figure*}
%%%%%%%%%%%%%%%%%%%%%%%%%%%%%%%%%%%%%%%%%%%%%%%%%%%%%%%%%%%%%%%%%%%%%%%%%%%%%%

\begin{lemma}
\label{lem:constraint_profile}
Let $\Tc$ be an optimal tree, and let $q^j$ and $q^{2k-2-j'}$ denote,
respectively, the heaviest and the lightest weights on level $M$ of the
tree. Then, we have $j\le k-1$, $j'\le k-1$, and $j+j' \le k$.
\end{lemma}

\begin{IEEEproof}
Consider first the case where $c > \ccmin$, i.e., all the components of
the profile $\nT[{\Tc}]$ are positive. The lightest weight on level
$M-1$ of the tree immediately precedes $q^j$ in $\bldp$. Hence, it is
of the form $q^{j-\eps}$, with $\eps\in\{0,1\}$. On the other hand,
reasoning similarly, the heaviest two weights on level $M+1$ are of the
form $q^{2k-2-j'+\eps'}$ and $q^{2k-2-j'+\eps'+\eps''}$, where
$\eps',\eps''\in\{0,1\}$ and $\eps'+\eps''\le 1$ (due to the ``three
consecutive weights'' property). Since $\Tc$ is optimal, by the
definition of $\DD$ in~(\ref{eq:defDc}), we must have $\DD\le 0$.
Applying~(\ref{eq:Dc}), the above constraints on $\eps,\eps',\eps''$,
and the fact that $q^k=\half$, we get
\begin{align*}
0 \ge \DD &= -q^{j-\epsilon}
  +q^{2k-2-j'+\epsilon'} +q^{2k-2-j'+\epsilon'+\epsilon''} \\
  &\ge
  -q^{j-1}+2 q^{2k-1-j'} = -q^{j-1}+ q^{k-1-j'}\,.
\end{align*}
Thus, $j+j'\le k$. Since both $j$ and $j'$ are positive when
$c>\ccmin$, the claim of the lemma follows in this case.

Consider now the case where $c=\ccmin$, i.e., $\Tc$ is a quasi-uniform
tree. If $\short=0$, we have $n_{M+1}=0$, and, thus, the lightest
weight on level $M$ is $p_{k^2}=q^{2k-2}$, and $j'=0$. For the heaviest
weight on level $M$, we have $p_{2^m-k^2+1} = q^j$.
By~(\ref{eq:local1}), we have $2^m-k^2+1 \le k(k+1)/2$. Recalling the
order and structure of $\bldp$, we obtain
$
q^j = p_{2^m-k^2+1} \ge p_{k(k+1)/2} = q^{k-1}\,.
$
Thus, $j\le k-1$. The case of $c=\ccmin$ and $\short=1$ is argued
similarly, using~(\ref{eq:local2}) in lieu of~(\ref{eq:local1}), and
leading to $j=0$ and $j'\le k-1$.
\end{IEEEproof}

It follows from Lemma~\ref{lem:constraint_profile} that in an optimal
tree, the heaviest weight on level $M$ is covered by~(\ref{eq:large})
in Lemma~\ref{lem:correspondence} (and, thus, so is any weight on level
$M-1$), while the lightest weight on level $M$ is covered
by~(\ref{eq:small}) in that lemma (and, thus, so is any weight on level
$M+1$). Consequently, an optimal tree is completely determined by a
tuple $\jvector=(j, r, j', r')$, with $0 \le j,j' \le k-1$, $ 0 \le r
\le j$, and $0 \le r' \le j'$. The profile of the tree is then given by
\begin{eqnarray}\squeeze
n_{M-1} & = & \frac{j(j+1)}{2}+r\,, \label{eq:nM-1jr}\\
n_{M+1} & = & \frac{j'(j'+1)}{2}+r'\,,\label{eq:nM+1j'r'}\\
n_M     & = & k^2 - n_{M-1} - n_{M+1}\,.\label{eq:nM}
\end{eqnarray}

The following lemma presents a characterization of the least value of
$c$ for which $\Tc$ is optimal. The lemma follows immediately from
Theorem~\ref{theo:choose} and Lemma~\ref{lem:dicho}.

\begin{lemma}\label{lem:ck with Dc}
Let $c_k$ be the least value of $c$ such that $\Tc$ is optimal. Then,
either $\DD[\ccmin+1]\ge 0$ (with $c_k=\ccmin$), or $\DD[c_k] < 0 $ and
$\DD[c_k+1]
\ge 0$ (with $c_k > \ccmin$).
\end{lemma}

Define the function
\begin{equation}\label{eq:Ffxn}
\Ffxn(\jtuple) = 2k^2-2^{M+1}+ j(j+1)+2r - \frac{j'(j'+1)}{2} - r'\,,
\end{equation}
acting on tuples $\jvector=(\jtuple)$ for a given value of $k$. Next,
we derive a set of conditions on the tuple $\jvector$ corresponding to
the tree $\Tck$ characterized in Lemma~\ref{lem:ck with Dc}.

\begin{lemma}\label{lem:c_star}
Let $\jvector=(\jtuple)$ be the tuple defining the profile of $\Tck$ in
(\ref{eq:nM-1jr})--(\ref{eq:nM}). Then,
\begin{equation}
\label{eq:jpp}
\Ffxn(\jtuple) = 0,
\end{equation}
and exactly one of the following conditions holds:
\begin{itemize}
\item[(i)] $j, j' > 0$,  $j+j'=k-2$. Either $r=0$ and $0\le r' \le j'$,
 or $1\le r \le j$ and $r'\in \{0, 1\}$.
\item[(ii)]
 $j, j' >0$,  $j+j'=k-1$, $r=0$ and $r'\in\{0, 1\}$.
\item[(iii)] $j'=0$, $r'=0$, $j \in \{k-2, k-1\}$, $0 \le r \le j$.
\item[(iv)] $j=0$, $r=0$, $j' \in \{k-2, k-1\}$, $0 \le r' \le j'$.
\end{itemize}
Conversely, if $\jvector=(\jtuple)$ satisfies~(\ref{eq:jpp}) and one of
the conditions~(i)--(iv), then $\jvector$ defines $\Tck$.
\end{lemma}

\begin{IEEEproof}
The necessity of~(\ref{eq:jpp}) follows from the definition of
$\Ffxn(\jtuple)$ and from~(\ref{eq:nm1B}), setting $c=\half n_{M+1}$,
substituting the expressions from~(\ref{eq:nM-1jr})
and~(\ref{eq:nM+1j'r'}) for $n_{M-1}$ and $n_{M+1}$, respectively, and
rearranging terms. In fact,~(\ref{eq:jpp}) must hold for any optimal
tree, not just for $c=\cck$. Conditions (i)--(iv) will follow from an
exhaustive case study of configurations that yield the inequalities on
the quantities $\DD$ that characterize the point $c=\cck$, as stated in
Lemma~\ref{lem:ck with Dc}.

Consider, first, the case where $\cck > \ccmin$. Then, for $c=\cck$, by
Lemma~\ref{lem:ck with Dc}, we have $\DD < 0$ and $\DD[c+1]\ge 0$.
Writing down the expressions for $\DD$ and $\DD[{c+1}]$ explicitly
according to~(\ref{eq:Dc}), we observe that six weights are involved,
as illustrated in Figure~\ref{fig:sixweights}. In order to switch from
a negative $\DD$ to a nonnegative $\DD[{c+1}]$, we must have a decrease
from $p_{2^M-k^2+c}$ to $p_{2^M-k^2+c+1}$, or an increase from
$p_{k^2-2c+1}+p_{k^2-2c+2}$ to $p_{k^2-2c-1}+p_{k^2-2c}$, or both. By
the definitions of $j$ and $j'$, we have $p_{2^M-k^2+c+1}=q^j$, and
$p_{k^2-2c}=q^{2k-2-j'}$. Taking into account that consecutive weights
can vary at most by a factor of $q$, we can write, for the other
weights involved,
\begin{align}
p_{2^M-k^2+c} & =  q^{j-\eps},\label{eq:eps1}\\
p_{k^2-2c-1}  & =  q^{2k-2-j'-\eps'},\\
 p_{k^2-2c+1} & =  q^{2k-2-j'+\eps''},\\
p_{k^2-2c+2}  & =   q^{2k-2-j'+\eps''+\eps'''}\,,\label{eq:eps2}
\end{align}
where $\epstuple\in\{0,1\}$, and, due to the ``three consecutive
weights'' property, we must have $\eps'+\eps''\le 1$ and $\eps'' +
\eps''' \le 1$. Table~\ref{tab:epsilons} summarizes the patterns of
values of $\bldeps=(\epstuple)$ that satisfy these constraints and also
produce the combination of weight increases or decreases necessary to
satisfy the conditions for $c=c_k$. On the right column of the table,
we list the conditions imposed on $\jvector$ by the constraints of each
case. To illustrate the proof approach, we derive these conditions,
below, for the representative case $\bldeps=(1,0,0,1)$. The other cases
follow using similar arguments, which are also similar to those used in
the proof of Lemma~\ref{lem:constraint_profile} (here, more parameters
are assumed known, which allows us to obtain tighter bounds).

Assume $\bldeps=(1,0,0,1)$. Then, writing the conditions on $\DD$ and
$\DD[c+1]$ at $c=\cck$ explicitly, substituting for the weights using
the known values in $\bldeps$, and recalling that $q^k=\half$, we
obtain
\begin{align*}
0 &> \DD = p_{k^2-2c+1} + p_{k^2-2c+2} - p_{2^M-k^2+c} \\
  & = q^{2k-2-j'} +
q^{2k-1-j'}- q^{j-1} \\
&> 2q^{2k-1-j'}-q^{j-1} = q^{k-1-j'}-q^{j-1}\,,%\label{eq:epsDc}
\end{align*}
and
\begin{align*}
0 &\le \DD[c+1] = p_{k^2-2c-1} + p_{k^2-2c} - p_{2^M-k^2+c+1} \\
  &= q^{2k-2-j'} +
q^{2k-2-j'}- q^{j} \\
&= 2q^{2k-2-j'}-q^{j} = q^{k-2-j'}-q^{j}\,.%\label{eq:epsDc+1}
\end{align*}
It follows that $k-2 \le j+j' \le k-1$, as claimed in the second row of
Table~\ref{tab:epsilons}. The conditions on $r$ and $r'$ follow from
Lemma~\ref{lem:correspondence}, observing that $r$ resets to zero at
points where $j$ increases, and similarly with $r'$ relative to $j'$.
In this case, $p_{2^M-k^2+c}$ is the last weight of the form $q^{j-1}$,
and, thus, we have $n_{M-1} = 2^M-k^2+c = j(j+1)/2$ and $r=0$; scanning
$\bldp$ from right to left, $p_{k^2-2c+2}$ is the last weight of the
form $q^{2k-1-j'}$, and, thus, we have $n_{M+1} = 2c = j'(j'+1)/2+1$,
and $r'=1$.

%%%%%%%%%%%%%%%%%%%%%%%%%%%%%%%%%%%%%%%%%%%%%%%%%%%%%%%%%%%%%%%
\begin{table}
\caption{\label{tab:epsilons} The possible cases for
$\bldeps=(\epstuple)$ from
  (\ref{eq:eps1})--(\ref{eq:eps2}), and the conditions imposed
  on $(\jtuple)$ at $c=c_k$.}
\vspace{-1.5em}
\begin{center}
\begin{small}
\renewcommand{\arraystretch}{1.20}
\begin{tabular}{|l|l|}
\hline
  $(\epstuple)$ & Conditions on $(\jtuple)$ \\
 \hline
(1,0,0,0) & $j+j'=k-2$, $r=0$, $2 \le r'\le j'-1$\\
  \hline
(1,0,0,1) & $j+j'\in\{k-2, k-1\}$, $r=0$, $r'=1$\\
  \hline
(1,0,1,0) &$j+j'\in\{k-2, k-1\}$, $r=0$, $r'=0$\\
  \hline
(0,0,0,1) & $j+j'=k-2$, $1\le r \le j$, $r'=1$ \\
  \hline
(0,0,1,0) & $j+j'=k-2$, $1\le r \le j, \quad r'=0$\\
  \hline
(1,1,0,0) & $j+j'=k-2$, $r=0$, $r'=j'$\\
  \hline
(1,1,0,1) & $j+j' \in\{k-1, k-2\}$, $r=0$, $r'=j'=1$\\
  \hline
(0,1,0,0) & case cannot occur at $c=\cck$\\
  \hline
(0,1,0,1) & $j+j'=k-2$, $1 \le r \le j$, $r'=j'=1$\\
\hline
\end{tabular}
\end{small}
\end{center}
\end{table}

It is readily verified that all the cases on the right column of
Table~\ref{tab:epsilons} satisfy either Condition~(i) or Condition~(ii)
of the lemma.

Consider now the case where $\cck=\ccmin$. In this case, the tree is
quasi-uniform. When $\short_k=0$, since $n_{M+1}=0$, we have $j'=r'=0$.
The condition $j\le k-1$ was established in
Lemma~\ref{lem:constraint_profile}, while the condition $j\ge k-2$
follows directly from $\DD[\ccmin+1] = \DD[1]
\ge 0$. Thus, Condition~(iii) of the lemma is satisfied in this case.
Similarly, when $\cck=\ccmin$ and $\short_k=1$, we have $j=r=0$, $j'
\le k-1$ was established in Lemma~\ref{lem:constraint_profile},
and $j' \ge k-2$ follows from
$\DD[\ccmin] \ge 0$. Thus, Condition~(iv) of the lemma is satisfied in
this case.

To prove the sufficiency of the conditions of the lemma, we first claim
that, with $\jvector$ satisfying the conditions, the profile
$\nN=(n_{M-1},n_M,n_{M+1})$ defined in~(\ref{eq:nM-1jr})--(\ref{eq:nM})
defines a valid tree. Clearly, $n_{M-1}$ and $n_{M+1}$ are
non-negative. To verify that $n_M$ is also non-negative, we write
\begin{align*}
n_{M-1} + n_{M+1} &= \frac{j(j+1)}{2} + \frac{j'(j'+1)}{2} + r + r'\\
&< \frac{(j+j'+1)^2}{2}+j+j',
\end{align*}
where the inequality follows from the fact that $(a{+}b{+}1)^2 >
a(a{+}1){+}b(b{+}1)$ for $a,b\ge 0$, and from the inequalities $r\le j$
and $r'\le j'$. With $j{+}j'\le k{-}1$, it follows that
$n_{M-1}+n_{M+1} < k-1 + k^2/2 < k^2$. Hence, $n_M$, as defined
in~(\ref{eq:nM}), is positive. On the other hand,~(\ref{eq:jpp}),
together with the fact that the components of $\nN$ add up to $k^2$, is
equivalent to the Kraft equality for $\nN$. Therefore, $\nN$ defines a
valid tree $\Tc$. It is readily verified that if either Condition~(i)
or (ii) is satisfied, then the parameters $\pair$ of $\Tc$ satisfy $c
>\ccmin$, $\DD<0$, and $\DD[c+1]\le 0$. Thus, by Lemma~\ref{lem:cascade},
we have $c=\cck$. Similarly, if either Condition~(iii) or (iv) is
satisfied, we have $c=\ccmin$, $\DD[\ccmin+1]\ge 0$, and, again,
$c=\cck$.
\end{IEEEproof}

The following lemma explores some properties of the function
$\fxdelta(x)$ defined in~(\ref{eq:deltax}).

\begin{lemma}\label{lem:fxdelta(0,k)}
(i) For any $x$, we have $\fxdelta(x+1) = \fxdelta(x) + x + k\,$.

(ii) We have $\fxdelta(-1) \le 0$ and $\fxdelta(k) > 0$. Thus, $x_0$,
the largest real root of $\fxdelta$, satisfies $-1 \le x_0 < k$.

(iii) The values $\fxdelta(k-1)$ and $\fxdelta(k-2)$ are even integers.
\end{lemma}
\begin{IEEEproof}
(i) The claim is readily verified by direct
application of~(\ref{eq:deltax}).

(ii) Setting $x=-1$ in~(\ref{eq:deltax}), and recalling that $\Qk
=k^2-\lceil {k(k-1)}/{4}\rceil$ and $M=\lceil \log Q\rceil$, we obtain
\begin{align*}
  \fxdelta(-1) &=2(k^2-\frac{k(k{-}1)}{4}-2^M)\\
  &= 2\big(\Qk-2^M+\frac{1}{2} \iverson{(k \bmod 4) \in\{ 2,
3\}}\big)\\
  & = \iverson{(k \bmod 4) \in\{ 2,
3\}} +2 (\Qk{-}2^M),
\end{align*}
where $\iverson{\predicate}=1$ if the predicate $\predicate$ is true,
or $\iverson{\predicate}=0$ otherwise. It follows that $\fxdelta(-1)$
can be positive only if $(k \bmod 4) \in\{ 2, 3\}$ and $\Qk=2^M$.
Writing $\Qk=\Qk(k)$, and computing explicitly $ Q(4\ell+2) =
(4\ell+3)(3\ell+1)$ and $Q(4\ell+3) = (\ell+1)(12\ell+7)$, we conclude
that $\Qk$ has at least one odd divisor when $(k \bmod 4)\in
\{2,3\}$. Therefore, we must have $\fxdelta(-1) \le 0$.

Furthermore, since $\Qk\le 2^M \le 2\Qk-1$, we have
\begin{align*}
\fxdelta(k)&=2k^2-2^{M+1}+k(k+1)-1\\
&\ge 2k^2-4\Qk+k(k+1)+1\\
&=-2k^2+4\left\lceil \frac{k(k-1)}{4}\right\rceil +k(k+1)+1\\
&\ge -2k^2+ k(k-1) +k(k+1)+1 =1.
\end{align*}
Thus, $\fxdelta(k) > 0$, and, since the coefficient of $x^2$ in
$\fxdelta(x)$ is $\half$, $x_0$ must be in the claimed range.

(iii) By direct computation, we have $\fxdelta(k-1)=
2\,{k}^{2}-{2}^{M+1}+ ( k-1 ) k$ and $\fxdelta(k-2)
=2\,{k}^{2}-{2}^{M+1}+ ( k-2)(k-1)$. Since $k>2$ and $M>0$, both values
are even.
\end{IEEEproof}

 To complete the proof of Theorem~\ref{theo:top}, we will construct
a tuple $\jvector=(\jtuple)$ that satisfies the conditions of
Lemma~\ref{lem:c_star}, and, thus, defines the sought parameter pair
$\pairk$.

\begin{IEEEproof}[Proof of Theorem~\ref{theo:top}]
It follows immediately from the definition of $\fxdelta(x)$
in~(\ref{eq:deltax}) and of $\Ffxn(\jtuple)$ in~(\ref{eq:Ffxn}) that
for $\jtuple$ we have
\begin{align*} %label{eq:Delta+F}
\Ffxn&(\jtuple) \\
&=\fxdelta(j) + \frac{(k-j-2)(k-j-1)}{2} - \frac{j'(j'+1)}{2} + 2r - r'\,.
\end{align*}
When $j' = k-j-2$, this reduces to
\begin{equation}\label{eq:Delta+F(k-2)}
\Ffxn(\jtuple) = \fxdelta(j) + 2r - r'\,,
\end{equation}
while with $j' = k-1-j$ we get
\begin{equation}\label{eq:Delta+F(k-1)}
\Ffxn(\jtuple) = \fxdelta(j) + 2r - r' - (k-j-1)\,.
\end{equation}
We will use these relations to verify that the solutions constructed
below satisfy~(\ref{eq:jpp}). Let $x_0$ be the largest real root of
$\fxdelta(x)$, and let $\xi =
\lfloor x_0 \rfloor$. By Lemma~\ref{lem:fxdelta(0,k)}(ii), we have $-1\le
\xi < k$, $\fxdelta(\xi) \le 0$, and $\fxdelta(\xi+1) > 0$.
We consider three main cases for $\fxdelta(\xi)$, and for each case
(and possible sub-cases) we define a tuple $\jvector=(\jtuple)$ and
verify that it satisfies the conditions of Lemma~\ref{lem:c_star}.
\begin{enumerate}
\item\label{fxdelta:1}
 $0 \le -\fxdelta(\xi) \le 2\xi\,$:\ \ Let $j=\xi$ , $r=\lfloor
 \frac{-\fxdelta(j)+1}{2}\rfloor$ and $r'= -\fxdelta(j)
 \mod 2$. By the assumptions of the case on $\fxdelta(\xi)$,
 we have $j\ge 0$. As for $j'$, we have the sub-cases below. At the
 end of each sub-case, we note which of Conditions~(i)--(iv) of
 Lemma~\ref{lem:c_star} is satisfied.
 \begin{enumerate}
 \item $j=0\,$:\ \ We must have $\fxdelta(0)=0$, so we get $r = r' =
 0$, and we set $j'=k-2$ (Condition~(iv)).
 \item $j\in \{k-2,k-1\}\,$:\ \ By Lemma~\ref{lem:fxdelta(0,k)}(iii),
$\fxdelta(j)$ is even, and $r'=0$. We get
  $r=-\frac{\fxdelta(j)}{2}$ and $0  \le r \le j$ by the
  assumptions on $\fxdelta(\xi)$, and we set $j'=0$ (Condition
  (iii)).
\item
$0 < j < k-2\,$:\ \ Set $j' = k - 2 - j$. From the choices for
$r$ and $r'$, we get $0 \le r \le j$ and $0 \le r' \le 1 \le
j'$ (Condition (i)).
\end{enumerate}
To verify that~(\ref{eq:jpp}) is satisfied, we
apply~(\ref{eq:Delta+F(k-2)}) for sub-cases a) and c), and for
sub-case b) with $j=k-2$. We apply~(\ref{eq:Delta+F(k-1)}) for
sub-case b) with $j=k-1$. For example, for sub-case c),
by~(\ref{eq:Delta+F(k-2)}) and the definitions of $r$ and $r'$, we
have,
\begin{align*}
\Ffxn(\jtuple)&= \fxdelta(j){+}2r{-}r'\\
& = \fxdelta(j){+}2\left\lfloor\frac{1{-}\fxdelta(j)}{2}\right\rfloor{-}r'\\
& =\fxdelta(j){+}2\frac{r'{-}\fxdelta(j)}{2}{-}r' = 0\,.
\end{align*}
Verification of $\Ffxn=0$ for the other sub-cases follows along
similar lines.

\item\label{fxdelta:2}
$-\fxdelta(\xi)\in\{2\xi+1,\,2\xi+2\}\,$:\ \ Let $j=\xi+1$. By
  Lemma~\ref{lem:fxdelta(0,k)}(ii), we have $0 \le j \le k$. We
  claim that $j\le k-1$. Assume, contrary to the claim, that $j=k$.
  Then, $-\fxdelta(k-1) = -\fxdelta(\xi) = 2k - \eps$ with
  $\eps\in\{0,1\}$, and, by Lemma~\ref{lem:fxdelta(0,k)}(i), we
  have $\fxdelta(\xi+1) = \fxdelta(k) = \fxdelta(k-1) + 2k-1 =
  \eps-1
  \le 0$, contradicting Lemma~\ref{lem:fxdelta(0,k)}(ii),
  which establishes $\fxdelta(\xi+1) > 0$. Thus, we have $0 \le j
  \le k-1$, and, defining $j' = k-1-j$, we also have $0 \le j' \le
  k-1$. By Lemma~\ref{lem:fxdelta(0,k)}(i), we have $\fxdelta(j) =
  \fxdelta(\xi+1) = \fxdelta(\xi) + \xi + k$, and, by the conditions
  of the case on $\fxdelta(\xi)$, we get $\fxdelta(j) \in \{k-j,
  k-j-1\}$. Define $r=0$, and $r' =
  \fxdelta(j) - (k-j-1)$, which implies $r' \in \{0,1\}$. Thus,
  whenever $0<j<k-1$, $\jvector=(\jtuple)$ satisfies Condition~(ii)
  of Lemma~\ref{lem:c_star}. When $j=0$, $\,\jvector$ satisfies
  Condition~(iv), and when $j=k-1$, it satisfies Condition~(iii) as
  long as $r'=0$. We claim that when $r'=1$, we must have $j<k-1$.
  Otherwise, if $r'=1$ and $j=k-1$, then, by the definition of
  $r'$, we have $\fxdelta(k-1)=\fxdelta(j)= r'+(k-j-1)= 1$,
  contradicting Lemma~\ref{lem:fxdelta(0,k)}(iii). Thus, $\jvector$
  satisfies one of the conditions (ii)--(iv) of
  Lemma~\ref{lem:c_star}. By~(\ref{eq:Delta+F(k-1)}) and the
  definitions of $r$ and $r'$, $\jvector$ also
  satisfies~(\ref{eq:jpp}).

\item\label{fxdelta:3}
$-\fxdelta(\xi) \ge 2\xi + 3\,$:\ \ Let $j=\xi+1$. By
  Lemma~\ref{lem:fxdelta(0,k)}(ii), we have $0 \le j \le k$. We
  claim that $j\le k-2$. Assume, contrary to the claim, that
  $j=k-1$. Then, $\xi=k-2$, and, by the assumptions of the case, we
  have $-\fxdelta(k-2) \ge 2(k-2)+3 = 2k-1$. Applying
  Lemma~\ref{lem:fxdelta(0,k)}(i), we get $\fxdelta(\xi+1) =
  \fxdelta(k-1) =
  \fxdelta(k-2) + (k-2) + k = \fxdelta(k-2) + 2k-2 \le -1$,
  contradicting Lemma~\ref{lem:fxdelta(0,k)}(ii), since we must
  have $\fxdelta(\xi+1) > 0$. Similarly, if $j = k$, then
  $-\fxdelta(k-1) \ge 2k+1 $ and $\fxdelta(k) = \fxdelta(k-1)+2k-1
  \le -2$, again contradicting Lemma~\ref{lem:fxdelta(0,k)}(ii). Thus,
  we have $ 0 \le j \le k-2 $, and we can define $j' = k-2-j$,
  which also satisfies $ 0 \le j' \le k-2$. By
 Lemma~\ref{lem:fxdelta(0,k)}(i), and the conditions of
  the case on $\fxdelta(\xi)$, we have $\fxdelta(j) =
  \fxdelta(\xi+1) = \fxdelta(\xi) + \xi + k \le k - \xi - 3 = k - 2 - j
  = j'$. Define $r=0$, and $r' = \fxdelta(j)$, satisfying $0 \le r'
  \le j'$. Thus, $\jvector=(\jtuple)$  satisfies Condition~(i)
  of Lemma~\ref{lem:c_star}. By~(\ref{eq:Delta+F(k-2)}) and the
  definitions of $r$ and $r'$, $\jvector$ also
  satisfies~(\ref{eq:jpp}).
\end{enumerate}

Cases~\ref{fxdelta:1}--\ref{fxdelta:3} above cover all possible values
of $\fxdelta(\xi)$, and in all cases, we have exhibited an explicit
tuple $\jvector=(\jtuple)$ satisfying the conditions of
Lemma~\ref{lem:c_star}, and, therefore, defining the optimal tree
$\Tskck$. It can readily be verified that the definitions of $j$ and
$r$ in~(\ref{eq:jrk1}) summarize the corresponding
definitions in the cases of the proof, with the top branch of~(\ref{eq:jrk1})
corresponding to Case~\ref{fxdelta:1}, and the bottom branch to
Cases~\ref{fxdelta:2} and~\ref{fxdelta:3}. Furthermore, the definition
of $c_k$ in~(\ref{eq:c}) reflects the parameter $c =n_{M-1}-2^M+k^2$ in
the profile~(\ref{eq:nM-1jr})--(\ref{eq:nM}) defined by $\jvector$ for
$c=c_k$.
\end{IEEEproof}

\begin{IEEEproof}[Proof of Corollary~\ref{cor:golomb-not-optimal}]
By the structure of $\Cplus$ in Theorem~\ref{theo:opt-code}, it
suffices to prove that $Q_k \concat Q_k$ is not optimal for the finite
source ${\mathcal A}_k$. Let $h=\lceil \log k \rceil$ and $a=2^h-k$,
with $0 \le a < 2^{h-1}$.  From the profile of $Q_k$ given in in
Section~\ref{sec:basics}, one derives the profile of $Q_k\concat Q_k$,
obtaining
\[ \nT[{Q_k\concat Q_k}]{=}
\big( n_{2h-2},\,n_{2h-1},\,n_{2h} \big)
{=}\left( a^2,\, 2a(k{-}a),\,(k{-}a)^2 \right)\,.
\]
Since $Q_k \concat Q_k$ has fringe thickness $\fT\le 2$, it has a
representation $\Tsgcg$, for some parameters $\short_g,\,c_g$, as
defined in Lemma~\ref{prop:profiles1}, with $N=k^2$. The case $a=0$
(i.e., $k=2^h$) is readily discarded as sub-optimal for $k>2$, as it
corresponds to a uniform tree with $2^{2h}$ leaves, which cannot be
optimal for $\Ak$ since $p_{k^2}+p_{k^2-1}<p_1$ for that source. Also,
we can assume that $\short_g$ is such that Lemma~\ref{lem:dicho} is
satisfied, and that $n_{2h-2}$ and $n_{2h}$ are such that they can be
written, respectively, as $n_{M-1}$ and $n_{M+1}$
in~(\ref{eq:nM-1jr})--(\ref{eq:nM+1j'r'}), with $j$ and $j'$ satisfying
Lemma~\ref{lem:constraint_profile}. Otherwise, $\Tsgcg$ is not optimal,
and the corollary is proved. By Lemma~\ref{lem:correspondence}, we can
write $a^2 < \half(j+1)(j+2) <
\half(j+2)^2$, or $j > \sqrt{2}\,a - 2$. Similarly, we have $(k-a)^2  <
\half(j'+1)(j'+2) <
\half(j'+2)^2$, or $j' > \sqrt{2}(k-a)-2$. Adding up, we obtain
$j+j' > \sqrt{2}\,k-4$, and, hence, for $k \ge 10$, $j+j' > k $,
contradicting Lemma~\ref{lem:constraint_profile}. For the remaining
cases, if $k\in\{7,9\}$ one verifies that $\short_g$ violates
Lemma~\ref{lem:dicho}, and for $k\in\{3,5,6\}$, one can easily verify,
by direct inspection, that $\Tsgcg$ is sub-optimal for $\Ak$.
\end{IEEEproof}

\section{Proofs for Subsection~\ref{sec:averagecodelength}}
\label{app:averagecodelength}

\begin{IEEEproof}[Proof of Corollary~\ref{cor:averagelength}]
By Theorem~\ref{theo:opt-code}, the code length for $(a,b)$ under
$\Cplus$ is $\len{T_k(a \bmod k,b \bmod k)} + 2 +
\lfloor \frac{a}{k}\rfloor + \lfloor \frac{b}{k}\rfloor\,$.
Writing $a=mk+i$ and $b=nk+j$ with $0\leq i, j< k$, $m,n\ge 0$, the
average code length under $\Cplus$ is
\begin{align}
\ncostt_q&(\Cplus) \nonumber\\
&=(1{-}q^2)\!\!\sum_{0 \le i,j < k}\!\!\!\!\!\sum_{\;\;\;\;m,n\ge 0} \!\!q^{i+j+(m+n)k}%
\bigl(
\len{T_k(i,j)}{+}m{+}n{+}2\bigr)\nonumber\\
&=\frac{2}{1-q^k}+ \frac{(1-q)^2}{(1-q^k)^2}\,\sum_{0\le
i, j\le k-1} \len{T_k(i,j)} q^{i+j}\nonumber\\
&= \frac{2}{1-q^k} + \ncost{T_k}\,,\label{eq:codelength2}
\end{align}
where the second equality follows from elementary series computations,
and the third identifies the (normalized) average code length of the
code $T_k$ defined in Theorem~\ref{theo:top}. Denote by $W_{M-1}, W_M$,
and $W_{M+1}$ the total normalized weight of symbols in $\Ak$ assigned
length $M-1,\,M$, and $M+1$, respectively, by $T_k$. Then, the average
code length of $T_k$ is given by
\begin{align}\label{eq:costTk}
\ncost{T_k} & = (M-1)\,W_{M-1} + M \, W_M + (M+1) W_{M+1} \nonumber\\
& = M +
W_{M+1}-W_{M-1}\,.
\end{align}
From the profile~(\ref{eq:cprofile}), with $N=k^2$ and $c=c_k$ as
defined in~(\ref{eq:c}), recalling~(\ref{eq:large}), letting $\gamma =
(1-q)^2/(1-q^k)^2$, and carrying out the computations, we obtain
\begin{align*}
W_{M-1}& = \gamma\sum_{i=1}^{j(j+1)/2+r} p_i =
\gamma\sum_{\ell=0}^{j-1} (\ell+1)q^{\ell} + \gamma\,r\,q^j\\
&= \frac{1-q^j\left(1+(1-q)j-(1-q)^2 r\right)}{(1-q^k)^2}\,.
\end{align*}
Similarly, from the proof of Theorem~\ref{theo:top}, setting $j'=k-j-2$
and $r'=2r+\fxdelta(j)$, we obtain
\begin{align*}
& W_{M+1}   = \gamma\sum_{i=0}^{j'(j'+1)/2+r'-1} p_{k^2-i} \\
& =
\gamma\sum_{\ell=0}^{j'-1}(\ell+1)
q^{2k-2-\ell}+\gamma\,r'\,q^{2k-2-j'}\\
& ={\frac {{q}^{2\,k}{+}{q}^{k+j} \Bigl( (k{-}j{-}1)(1{-}q)q-q
+ ( 1{-}q )^{2} \bigl( 2r{-}\fxdelta(j)  \bigr)\Bigr)}
{ \left( 1-{q}^{k} \right)
^{2}}}\;.
\end{align*}
The result~(\ref{eq:avglenCk2}) now follows by substituting the above
expressions for $W_{M-1}$ and $W_{M+1}$ in~(\ref{eq:costTk}),
substituting for $\ncost{T_k}$  in~(\ref{eq:codelength2}), and using
appropriate algebraic simplifications. The
result~(\ref{eq:avglenCkqk}), in turn, follows by applying the relation
$q^k = 1/2$.
\end{IEEEproof}

\section{Layer transitions in the codes $\Cminus$}
\label{app:layers}
In each layer transition described below, we assume that we start from
a layer $\layer$ of type (x), and show how it unfolds into a layer
$\layer[s+1]$ of type (y), the transition being denoted (x)$\to$(y). We
denote by $d_s$ the depth of the shallowest node in $\layer$.

\begin{description}
\labelwidth 8ex
\labelsep 1ex
\itemindent 0ex
\itemsep 0ex
\item[(i)$\to$(i):\ ]
The tree $q^{s+1} \Ltreek$ in each of the $\ell$ groups
$\macrosymbol$ in $\layer$ unfolds, by the definition of $\Ltreek$
(see also Figure~\ref{fig:ltrees}), into a tree $q^{s+2} \Ltreek$
and $2^{k}-1$ leaves of weight $q^{s+1}$, which provides a group
$\macrosymbol$ for $\layer[s+1]$. Hence, there are $\ell$ groups
$\macrosymbol$ in $\layer[s+1]$, which include $(2^k-1)\ell$
signatures $s+1$. This propagation of groups $\macrosymbol$ will
occur in the same way in all the other transitions below; its
discussion will be omitted for those cases. There remain
$s+2-(2^k-1)\ell = 2^{k-1}+1+j$ signatures $s+1$, with $0 \le j\le
2^{k-1}-4$ (recall that layers of type (i) exist only if $k>2$). A
quasi-uniform tree with $2^{k-1}+2+j$ leaves is built, rooted at
$\rsymb$. This tree has $2^{k-1}-(j+1)-1$ leaves at depth $k-1$,
which are labeled $s+1$, and $2(j+1)+2$ leaves at depth $k$, of
which $2(j+1)+1$ are assigned label $s+1$, and one serves as the
root of $\rsymb[s+1]$, consistent with a structure of type (i) for
$s+1$ (and, correspondingly, $j+1$).

\item[(i)$\to$(ii):\ ]
We have $j=2^{k-1}-3$. We let $\rsymb$ be the root of a balanced
tree of height $k$. Of its $2^k$ leaves, $2^k-2$ are assigned the
remaining $2^k-2$ signatures $s+1$, one leaf serves as the root for
$q\, \Ctree_{k-1}$, and the remaining leaf as the root for
$\rsymb[s+1]$.

\item[(ii)$\to$(iii) ($k{>}2$):\ ]
\hspace{2.75em} The tree $q\,\Ctree_{k-1}$ in $\layer$ contributes $2^{k-1}$ leaves
 of signature $s+1$ to $\layer[s+1]$, in addition to those
 contributed by the groups $\macrosymbol$. There remain $2^{k-1}-1$
signatures $s+1$, which are assigned to leaves of a balanced tree
 $\Ctree_{k-1}$ rooted at $\rsymb$. The remaining leaf splits into
two nodes, one is the root of a tree $q\,\Ctree_{k-1}$, and the
other anchors $\rsymb[s+1]$.

\item[(ii)$\to$(iv) ($k{=}2$):\ ]
\hspace{2.75em} The tree
$q\,\Ctree_{1}$ in $\layer$ contributes $2^1$ leaves of
signature $s+1$ to $\layer[s+1]$, in addition to those
contributed by the groups $\macrosymbol$. The remaining
signature $s+1$ is assigned to one leaf of a tree $\Ctree_1$
rooted at $\rsymb$. The second leaf splits into two
 nodes, one is the root of a tree $q\Ltreem$, and the other
 anchors $\rsymb[s+1]$.

\item[(iii)$\to$(iii):\ ]
\ \ The construction from the previous transition is kept, except that
one of the leaves of the tree $\Ctree_{k-1}$ rooted at $\rsymb$ is
split, making room for the additional signature $s+1$ resulting
from the increase in $s$. Hence, there is a decrease by one in the
number of leaves at depth $d_s$ and an increase by two in the
number of leaves at depth $d_s+1$. This process continues until
$j=2^k-4$.

\item[(iii)$\to$(iv):\ ]
\hspace{0.1em} This transition is identical to the previous one, except that
instead of a tree $q\,\Ctree_{k-1}$, a tree $q\Ltreem$ is attached
as sibling to $\rsymb[s+1]$.

\item[(iv)$\to$(v):\ ]
\hspace{0.1em} The tree $q\Ltreem$ from the previous transition provides the
$2^{k-1}-1$ leaves of signature $s+1$, plus a tree $q\Ltreek$. What
started as a balanced tree of depth $k-1$ in the transition
(ii)$\to$(iii) has evolved into a balanced tree of depth $k$, with
all leaves assigned signatures $s+1$, except for one, which serves
as the root of $\rsymb[s+1]$.

\item[(v)$\to$(i) ($k{>}2$):]
\hspace{2.2em} The tree $q\Ltreek$ added in the previous transition
generates a new group $\macrosymbol$, consistent with the increment
in $\ell$. All signatures $s+1$ now originate from the groups
$\macrosymbol$, or from $\rsymb$, which brings the construction
back to a layer of type (i), completing the cycle.

\item[(v)$\to$(ii) ($k{=}2$):]
\hspace{2.5em} When $k=2$ the transition occurs to a layer of type
(ii), as described above for the initial transition from Case 1 to
Case 2.
\end{description}

%\bibliography{2DGD}

\end{document}